\documentclass[12pt]{article}
\newcommand{\bfm}[1]{\mbox{\boldmath{$#1$}}}
\newcommand{\beq}{\begin{eqnarray}}
\newcommand{\eeq}{\end{eqnarray}}
\newcommand{\beqs}{\begin{eqnarray*}}
\newcommand{\eeqs}{\end{eqnarray*}}

\usepackage{graphics,epsfig}
\usepackage{latexsym}
\usepackage{amsthm}
\usepackage{amssymb,amsfonts,amsmath}

\newtheorem{theorem}{Theorem}
\newtheorem{corollary}{Corollary}
\newtheorem{hypothesis}{Hypothesis}

\begin{document}

\title{Minimum Energy Configurations in the $N$-Body Problem and the Celestial Mechanics of Granular Systems}
\author{D.J. Scheeres\\Department of Aerospace Engineering Sciences\\The University of Colorado at Boulder\\scheeres@colorado.edu}
\date{\today}
\maketitle

\begin{abstract}

Minimum energy configurations in celestial mechanics are investigated. It is shown that this is not a well defined problem for point-mass celestial mechanics but well-posed for finite density distributions. This naturally leads to a granular mechanics extension of usual celestial mechanics questions such as relative equilibria and stability. This paper specifically studies and finds all relative equilibria and minimum energy configurations for $N=1,2,3$ and develops hypotheses on the relative equilibria and minimum energy configurations for $N\gg 1$ bodies.

\end{abstract}

\section{Introduction}

Celestial Mechanics systems have two fundamental conservation principles that enable their deeper analysis: conservation of momentum and conservation of (mechanical) energy. Of the two, conservation of momentum provides the most constraints on a general system, with three translational symmetries (which can be trivially removed) and three rotational symmetries. If no external force acts on the system, these quantities are always conserved independent of the internal interactions of the system. Conservation of energy instead involves assumptions on both the lack of exogenous forces and on the nature of internal interactions within the system. For this reason energy is often not conserved for ``real'' systems that involve internal interactions, such as tidal deformations or impacts, even though they may conserve their total momentum. Thus mechanical energy generally decays through dissipation until the system has found a local or global minimum energy configuration that corresponds to its constant level of angular momentum.
This observation motivates a fundamental question for celestial mechanics: \\

{\centering\it What is the minimum energy configuration of a $N$-body system with a fixed level of angular momentum?}\\ 

This paper shows that this is an ill-defined question for traditional point-mass celestial mechanics systems. If instead the system and problem are formulated accounting for finite density distributions this question becomes well posed and provides new light on celestial mechanics systems. 

This is a well motivated adjustment as real systems always have a finite density and, hence, any particle in a celestial mechanics system has a finite size. Such a physically corrected system has been called the ``Full $N$-Body Problem,'' as inclusion of finite density also necessitates the modeling of the rotational motion of the components, which is not needed for consideration of point masses. It also necessitates consideration of contact forces as their mass centers cannot come arbitrarily close to each other, as at some distance they will rest on each other. Thus, introduction of this finite density correction allows the minimum energy configurations for an $N$-body system to be explicitly defined and computed for a given level of angular momentum. 

Given this perspective, an interesting problem is to track the absolute minimum energy configuration of a collection of $N$ particles as the system angular momentum increases from or decreases to zero. This is, essentially, an investigation of the celestial mechanics of granular systems as a function of total angular momentum. This problem has been shown to be relevant to the understanding of solar system bodies, especially among asteroids whose size is small enough so that when the components rest on each other they have insufficient gravitational attraction to overcome material strength, and thus retain the physical characteristics of rigid bodies resting on each other. Due to the celebrated YORP effect, which has established that non-symmetrically shaped bodies subject to solar radiation will change their spin rates over time, this question has several practical applications and has been implicated in how small asteroids form binary systems \cite{scheeres_fission}. 
The question of stable minimum energy configurations for the $N=2$ particle Full Body problem has been worked out in detail \cite{scheeres_gravgrad, scheeres_F2BP_planar}, and has been verified as a viable physical model via astronomical observations of asteroids \cite{pravec_fission}.

This paper presents a number of Theorems that motivate and enable our study of minimum energy configurations, and then explore some exact results for the cases of $N=1,2,3$. Some hypotheses for the case $N\gg 1$ are also given.  
\begin{description}
\item[Theorem 1:] A sharper version of the Sundman Inequality is derived: $H^2 \le 2 I_H T \le 2 I_p T$, where $H$ is the system angular momentum, $T$ is the system kinetic energy, $I_p$ is the system polar moment of inertia and $I_H$ is the system moment of inertia about a fixed direction in space. This form of the Sundman Inequality is the appropriate generalization for bodies with finite density distributions. 

\item[Theorem 2:] The minimum energy function ${\cal E}$ is defined: ${\cal E} = \frac{H^2}{2I_H} + U \le E$, where $U$ is the total potential energy of the system and $E$ is the total system energy. It is shown that ${\cal E} = E$ occurs if and only if the minimum energy function is stationary with respect to all variations for a given level of angular momentum $H$ and represents a relative equilibrium for the system (borrowing from a theorem originally proven by Smale).

\item[Theorem 3:] It is shown that the point-mass $N$-body problem with a specified level of angular momentum does not have either global or local minimum energy configurations for all $N\ge 3$ (borrowing from a theorem by Moeckel).

\item[Theorem 4:] It is shown that the finite-density $N$-body problem has a global minimum energy configuration for all $N\ge 1$ and for all levels of angular momentum.

\item[Theorem 5-8:] All minimum energy configurations, both local and global, are found for all levels of angular momentum for the finite density cases $N=1,2,3$. 

\item[Hypothesis 1:] It is hypothesized that for all $N$ the local and global minimum energy configurations for a finite density system will either condense into a single aggregate and spin at a uniform rate, or will lie in a two component system with the orbit and all rotations spinning at the same uniform rate. 

\end{description}

The current paper proves a number of fundamental results that motivate and enable the further exploration of these questions. 
These results show a surprising complexity in the evolution of minimum energy states as a function of angular momentum, with distinctly different pathways arising as the number of particles in the system increases. 

\section{Background}

\subsection{Conservation Principles}

A closed mechanical system has two fundamental conservation principles: conservation of momentum, both translational and rotational, and conservation of energy. The statement of these is simple and can be derived starting from the fundamental equations of motion for this system. This paper does not detail these results, as they are well known (see the explicit derivation of these conservation principles for the full 2-body problem in \cite{scheeres_F2BP}). 

The result is that the total translational momentum, angular momentum $\bfm{H}$ and total energy $E = T+U$ (where $T$ is kinetic energy and $U$ is gravitational potential energy) is conserved under motion within the system. By specifying the barycenter of the system to be at rest and at the origin of the system the three components of the translational momentum are removed, which allows the fundamental quantities of the system to be stated in a purely relative form. Angular momentum of the system is also conserved so long as no exogenous forces act -- regardless of the internal actions between the components. The same is not true of the energy, as its conservation depends on both the lack of exogenous forces and on strict specification of internal interactions. Specifically, all internal interactions must be reversible, meaning that no dissipation can occur. For ideal celestial mechanics involving point mass interactions this is a reasonable model -- however the point-mass celestial mechanics model is always an approximation due to its infinite density distribution at the particle location. The truth is that all physical celestial mechanics systems of interest involve dissipative interactions, and hence all natural systems will tend to dissipate energy, meaning that total energy is no longer conserved. In many cases the timescale for energy dissipation is quite slow, allowing for the system to be adequately modeled over extended periods of time as conserving energy. However, as has been well established in the study of the solar system, energy dissipation due to tidal distortion of bodies and internal stress wave propagation eventually dominates the dynamics and interactions of almost all celestial mechanics systems and forces them to evolve to lower energy states, all the while conserving total angular momentum \cite{goldreich_peale,goldreich}. 

Two simple examples of this can be mentioned. First is that almost all minor planets in the solar system are in or close to uniform rotation about their maximum moments of inertia -- this despite a past filled with mutual impacts and gravitational torques from their occasional interactions with planets. This is simply due to the fact that uniform rotation about the maximum moment of inertia is the minimum energy state for a solo rotating body. The actual dissipation occurs due to the interaction of time varying accelerations that the body experiences when not in uniform rotation with the material properties of the body, causing minute elastic deformations that turn mechanical energy into heat, which is then radiated away \cite{burns_safronov}. The other example is the mutual orbit of the Earth and moon. The moon has already shed its excess rotational energy and has settled into a local minimum energy state, with the same face always pointing at the Earth. The Earth still spins more rapidly than the mutual orbit, however, raising tides on the Earth which dissipate energy and slow the Earth's rotation rate. Conservation of angular momentum dictates that the mutual orbit expands -- this occurs physically by the torque that the Earth's tidal bulge places on the moon. The eventual state of the Earth-Moon system (ignoring the sun) is for both systems to devolve into mutually synchronous orbits, as is the case for the Pluto-Charon system. Once in this final state no excess energy dissipation can occur as the system is in its minimum energy state for its given total angular momentum. 

\subsection{Density Distributions and Fundamental Quantitites}

To discuss minimum energy configurations for the point mass or finite density $N$ body problem requires the definition of a few basic scalar and vector functions and quantities. Consider the total kinetic energy, gravitational potential energy, polar moment, and angular momentum of an arbitrary collection of $N$ mass distributions, denoted as ${\cal B}_i, i = 1,2, \ldots, N$. Each body ${\cal B}_i$ is defined by a differential mass distribution $dm_i$ that can fall into one of two forms, a point mass distribution or a finite density distribution, denoted as
\beq
	dm_i & = & \left\{ \begin{array}{cc} 
		m_i \delta(|\bfm{r}-\bfm{r}_i|) & \mbox{ Point Mass Density Distribution} \\
		\rho_i(\bfm{r}) \ dV & \mbox{ Finite Density Distribution} \end{array} \right. \label{eq:dmass}
\eeq
where $m_i$ is the total mass of body ${\cal B}_i$, $\delta(-)$ is the Dirac delta function, $|-|$ denotes the Euclidian norm, $\rho_i$ is the density of body ${\cal B}_i$ (possibly constant) and $dV$ represents the differential volume of the body. If ${\cal B}_i$ is described by a point mass density distribution, the body itself is just defined as a single point $\bfm{r}_i$. Instead, if the body is defined as a finite density distribution, ${\cal B}_i$ is defined as a compact set in $\mathbb{R}^3$ over which $\rho_i(\bfm{r}) \ne 0$. In either case the ${\cal B}_i$ are defined as compact sets. 

Assume that each differential mass element $dm_i(\bfm{r})$ has a specified position and  an associated velocity. For components within a given body ${\cal B}_i$ a rigid body assumption is made so that the entire body can be defined by the position and velocity of its center of mass, its attitude, and its angular velocity. Finally, assume that these positions and velocities are defined relative to the system barycenter, which is chosen as the origin, or
\beq
	\sum_{i=1}^N \int_{{\cal B}_i} \bfm{r} dm_i(\bfm{r}) & = & 0 \\
	\sum_{i=1}^N \int_{{\cal B}_i} \dot{\bfm{r}} dm_i(\bfm{r}) & = & 0 
\eeq

Given these definitions an integral form of the kinetic energy, polar moment, gravitational potential energy, and angular momentum vector can be stated as
\beq
	T & = & \frac{1}{2} \sum_{i=1}^N \int_{{\cal B}_i} \left( \dot{\bfm{r}}\cdot\dot{\bfm{r}} \right) dm_i(\bfm{r}) \\
	I_p & = &  \sum_{i=1}^N \int_{{\cal B}_i} \left( {\bfm{r}}\cdot{\bfm{r}} \right) dm_i(\bfm{r})  \label{eq:isum} \\
	U & = & - {\cal G} \sum_{i=1}^{N-1} \sum_{j=i+1}^N \int_{{\cal B}_i} \int_{{\cal B}_j} \frac{dm_i \ dm_j}{|\bfm{r}_{ij}|}  \label{eq:udef0} \\
	\bfm{H} & = & \sum_{i=1}^N \int_{{\cal B}_i} \left( {\bfm{r}}\times\dot{\bfm{r}} \right) dm_i(\bfm{r}) \label{eq:hsum}
\eeq
where $\bfm{r}_{ij} = \bfm{r}_j - \bfm{r}_i$. Note that the definition of $U$ in Eqn.\ \ref{eq:udef0} eliminates the self-potentials of these bodies from consideration. As the finite density mass distributions are rigid bodies this elimination is reasonable. 
This notation can be further generalized by defining the single and joint general mass differentials
\beq
	dm(\bfm{r}) & = & \sum_{i=1}^N   dm_i(\bfm{r}) \\
	dm(\bfm{r}) \ dm'(\bfm{r}') & = & \sum_{i=1}^{N-1} \sum_{j=i+1}^N     dm_i(\bfm{r}) \ dm'_j(\bfm{r}') 	
\eeq
and the total mass distribution ${\cal B} = \left\{ {\cal B}_i, i = 1, 2, \ldots, N\right\}$. Then the above definitions can be reduced to integrals over ${\cal B}$:
\beq
	T & = & \frac{1}{2} \int_{{\cal B}} \left( \dot{\bfm{r}}\cdot\dot{\bfm{r}} \right) dm(\bfm{r}) \\
	I_p & = &  \int_{{\cal B}} \left( {\bfm{r}}\cdot{\bfm{r}} \right) dm(\bfm{r}) \label{eq:ipint} \\
	U & = & - {\cal G} \int_{{\cal B}} \int_{{\cal B}} \frac{dm \ dm'}{|\bfm{r}-\bfm{r}'|}\\
	\bfm{H} & = & \int_{{\cal B}} \left( {\bfm{r}}\times\dot{\bfm{r}} \right) dm(\bfm{r}) \label{eq:hint} \\
	\int_{\cal B} \bfm{r} dm(\bfm{r}) & = & 0 \\
	\int_{\cal B} \dot{\bfm{r}} dm(\bfm{r}) & = & 0 
\eeq

\subsection{Point-Mass Results}

Assuming that all of the differential densities follow the point mass density distribution, the classical results for the $N$-body problem are recovered:
\beq
	T & = & \frac{1}{2} \sum_{i=1}^N m_i \left( \dot{\bfm{r}}_i\cdot\dot{\bfm{r}}_i \right) \\
	I_p & = &  \sum_{i=1}^N m_i \left( {\bfm{r}}_i\cdot{\bfm{r}}_i \right) \\
	U & = & - {\cal G} \sum_{i=1}^{N-1} \sum_{j=i+1}^N \frac{m_i \ m_j}{|\bfm{r}_{ij}|}\\
	\bfm{H} & = & \sum_{i=1}^N m_i \left( {\bfm{r}}_i\times\dot{\bfm{r}}_i \right) 
\eeq
It is convenient to apply Lagrange's Identity to the kinetic energy, polar moment of inertia and angular momentum, allowing them to be restated in a relative form. Note that the following relationships only hold if the barycenter of the system is at the origin.
\beq
	T & = & \frac{1}{2M} \sum_{i=1}^{N-1}\sum_{j=i+1}^N m_i m_j \left( \dot{\bfm{r}}_{ij}\cdot\dot{\bfm{r}}_{ij} \right) \\
	I_p & = & \frac{1}{M} \sum_{i=1}^{N-1}\sum_{j=i+1}^N m_i m_j \left( {\bfm{r}}_{ij}\cdot{\bfm{r}}_{ij} \right) \\
	\bfm{H} & = & \frac{1}{M} \sum_{i=1}^{N-1}\sum_{j=i+1}^N m_i m_j \left( {\bfm{r}}_{ij}\times\dot{\bfm{r}}_{ij} \right) 
\eeq

\subsection{Full-Body Results}

If finite density distributions are assumed for each body the point-mass results must be generalized to incorporate rotational kinetic energy, rigid body moments of inertia, angular velocities and explicit mutual potentials that are a function of body attitude\cite{scheeres_F2BP}. In the following the $i$th rigid body's center of mass is located by the position $\bfm{r}_i$ and has a velocity $\dot{\bfm{r}}_i$. In addition to its mass $m_i$, the $i$th body has an inertia dyadic $\bfm{I}_i$, an angular velocity vector $\bfm{\Omega}_i$ and an attitude dyadic that maps its body-fixed vectors into inertial space, $\bfm{A}_i$. The basic quantities are then defined as (generalizing results from \cite{scheeres_F2BP}):
\beq
	T & = & \frac{1}{2} \sum_{i=1}^N \left[ m_i \left( \dot{\bfm{r}}_i\cdot\dot{\bfm{r}}_i \right) + \bfm{\Omega}_i \cdot \bfm{I}_i \cdot \bfm{\Omega}_i \right] \\
	I_p & = &  \sum_{i=1}^N m_i \left[ \left( {\bfm{r}}_i\cdot{\bfm{r}}_i \right) + \frac{1}{2} \mbox{Trace}( \bfm{I}_{i} )\right] \\
	U & = & \sum_{i=1}^{N-1} \sum_{j=i+1}^N U_{ij}( \bfm{r}_{ij}, \bfm{A}_{ij} ) \label{eq:udef} \\
	\bfm{H} & = & \sum_{i=1}^N \left[ m_i \left( {\bfm{r}}_i\times\dot{\bfm{r}}_i \right) + \bfm{A}_i\cdot\bfm{I}_i\cdot\bfm{\Omega}_i \right]
\eeq
In the above the inertia dyadics are all specified in a body-fixed frame and thus are constant, the  $U_{ij}$ are mutual potentials between two different rigid bodies $i$ and $j$ and are only a function of their relative position and relative attitude, $\bfm{A}_{ij}$, equal to $\bfm{A}_j^T\cdot\bfm{A}_i$, and which transfers a vector from the body $i$ frame into the body $j$ frame. 

The kinetic energy, polar moment of inertia and angular momentum can again be stated in relative form between the center of masses, leaving the rotational components in their current form.
\beq
	T & = & \frac{1}{2M} \sum_{i=1}^{N-1}\sum_{j=i+1}^N m_i m_j \left( \dot{\bfm{r}}_{ij}\cdot\dot{\bfm{r}}_{ij} \right)  + \frac{1}{2} \sum_{i=1}^N \bfm{\Omega}_i \cdot \bfm{I}_i \cdot \bfm{\Omega}_i  \\
	I_p & = & \frac{1}{M} \sum_{i=1}^{N-1}\sum_{j=i+1}^N m_i m_j \left( {\bfm{r}}_{ij}\cdot{\bfm{r}}_{ij} \right) + \frac{1}{2} \sum_{i=1}^N \mbox{Trace}( \bfm{I}_i) \\
	\bfm{H} & = & \frac{1}{M} \sum_{i=1}^{N-1}\sum_{j=i+1}^N m_i m_j \left( {\bfm{r}}_{ij}\times\dot{\bfm{r}}_{ij} \right) + \sum_{i=1}^N \bfm{A}_i\cdot\bfm{I}_i\cdot\bfm{\Omega}_i
\eeq

\subsubsection{Alternate System Moment of Inertia}

An alternate version of the system moment of inertia can be defined once the direction of the total angular momentum is specified. First define the total inertia dyadic of the $N$ body system as
\beq
	\bfm{I} & = & - \int_{\cal B} \tilde{\bfm{r}}\cdot\tilde{\bfm{r}} dm
\eeq
where $\tilde{\bfm{r}}$ is the cross product dyadic defined such that $\bfm{a}\times\bfm{b} = \tilde{\bfm{a}}\cdot\bfm{b} = \bfm{a}\cdot\tilde{\bfm{b}}$. When evaluated in detail, and in a common inertial frame, the system inertia dyadic is equal to
\beq
	\bfm{I} & = & \sum_{i=1}^N \left[ m_i \left( r_{i}^2 \bfm{U} - \bfm{r}_{i}\bfm{r}_{i}\right) + \bfm{A}_i\cdot\bfm{I}_i\cdot\bfm{A}_i^T \right] 
\eeq
where the identity $- \tilde{\bfm{a}}\cdot\tilde{\bfm{b}} = (\bfm{a}\cdot\bfm{b}) \bfm{U} - \bfm{b}\bfm{a}$ is applied and where $\bfm{U}$ is the unity dyadic and the multiplication of two vectors is defined as a dyad. Given a defined direction of the total angular momentum vector in inertial space, $\hat{\bfm{H}}$, the moment of inertia relative to this direction can be defined as
\beq
	I_H & = & \hat{\bfm{H}}\cdot\bfm{I}\cdot\hat{\bfm{H}} \label{eq:ih} 
\eeq
This alternate version of the polar moment of inertia is of use in developing sharper limits for the Sundman Inequality later.
For the point mass density distributions all of the moments of inertia disappear and $I_p = I_H$ if the bodies and their velocities all lie in a common plane.

\subsubsection{Minimum Distances between Mass Collections}

A crucial aspect of the finite density distributions are that these bodies can come into contact with each other without singularities or deformations. Thus, under a rigid body assumption they will have minimum distances between their centers of mass as a function of their relative attitude. If the bodies are convex, strong results on the minimum distances between the bodies can be found. However, even complex shaped bodies will still have well defined minimum distances between the mass collections. 

Each body has a minimum diameter (equal to the diameter of the inscribing sphere and denoted by $d$) and a maximum diameter (equal to the diameter of the circumscribing sphere and denoted by $D$), these are equal only if the body is a sphere. The relative distance between any two mass distributions ${\cal B}_i$ and ${\cal B}_j$ with minimum diameters $d_i$ and $d_j$ is bounded from below by the average of these, or $r_{ij} \ge \frac{1}{2}(d_i + d_j) = d_{ij}$. For distances $r_{ij} > \frac{1}{2}(D_i + D_j) = D_{ij}$ the relative position and attitude between the two collections is unconstrained. For any combination of relative position $D_{ij}\hat{\bfm{r}}_{ij}$ and relative attitude $\bfm{A}_{ij}$ the distance between the two mass distributions can be decreased to their relative distance at which they will touch, defined as $d_{ij}(\hat{\bfm{r}}_{ij}, \bfm{A}_{ij})$. The contact points of bodies $i$ and $j$ that define this distance lie in their respective boundary sets of the mass distributions. For the given definition, the set of these minimum distances is a closed set and is bounded from below by $d_{ij}$, the minimum distance. For non-convex shapes these minimum distance sets can be complicated and may not even be able to attain the theoretical minimum. Also, in these cases different definitions of the minimum distance can be defined, depending on whether the bodies are star-convex or have a more convoluted shape. Note that across any of these shapes it is always possible to define a minimum bound on the distance between these shapes that can be realized, meaning that the distance between two mass distributions is a closed set along its lower bound. If the bodies are convex this follows trivially. 

\subsubsection{Finite Density Sphere Restriction}

This paper mainly focuses on the sphere-restriction of the Full-Body problem, where all of the bodies have finite, constant densities and spherical shapes defined by a diameter $d_i$. This allows for considerable simplification of the mutual potentials, although the rotational kinetic energy, moments of inertia and angular momentum of the systems are still tracked. In this case the moment of inertia of a constant density sphere is $m_i d_i^2 / 10$ about any axis and the minimum distance between two bodies will be $d_{ij} = (d_i + d_j)/2$. The resultant quantities for these systems are
\beq
	T & = & \frac{1}{2M} \sum_{i=1}^{N-1}\sum_{j=i+1}^N m_i m_j \left( \dot{\bfm{r}}_{ij}\cdot\dot{\bfm{r}}_{ij} \right)  + \frac{1}{2} \sum_{i=1}^N \frac{m_i d_i^2}{10} {\Omega}^2_i  \\
	U & = & - {\cal G} \sum_{i=1}^{N-1} \sum_{j=i+1}^N \frac{m_i \ m_j}{|\bfm{r}_{ij}|}  \\
	\bfm{H} & = & \frac{1}{M} \sum_{i=1}^{N-1}\sum_{j=i+1}^N m_i m_j \left( {\bfm{r}}_{ij}\times\dot{\bfm{r}}_{ij} \right) + \sum_{i=1}^N \frac{m_i d_i^2}{10} \bfm{\Omega}_i
\eeq
with the two different versions of the moment of inertia
\beq
	I_p & = & \frac{1}{M} \sum_{i=1}^{N-1}\sum_{j=i+1}^N m_i m_j \left( {\bfm{r}}_{ij}\cdot{\bfm{r}}_{ij} \right) + \frac{3}{2} \sum_{i=1}^N \frac{m_i d_i^2}{10}  \\
	I_{H} & = & \frac{1}{M} \sum_{i=1}^{N-1}\sum_{j=i+1}^N m_i m_j \left( r_{ij}^2 - (\hat{\bfm{H}}\cdot\bfm{r}_{ij})^2 \right) + \sum_{i=1}^N \frac{m_i d_i^2}{10}  
\eeq

\subsection{Energy Dissipation Interaction Models}

Implicit in these discussions, although not explicitly incorporated into the interaction models, are the dissipative effects of surface Coulomb friction and the tidal distortion of gravitationally attracting bodies. These physical effects serve dual purpose, in that they will tend to synchronize collections of bodies, either resting on each other or orbiting each other, and will also dissipate excess energy in the system. 

No surface interaction models between the finite density bodies is considered other than Coulomb friction between surfaces. This is needed in order for a resting collection of particles to dissipate relative motion between each other when in contact and thus also represents one possible mode of energy dissipation. Inclusion of this notional model ensures that contact configurations will, when reduced to their minimum energy state, all rotate at a common rate. It is possible to explicitly incorporate additional surface potentials between particles in close proximity, such as the Lennard-Jones potential that models van der Waals cohesive forces \cite{castellanos}. These extensions are not considered here, however.

It is also possible to dissipate energy and synchronize spin rates even if bodies are not in contact. Tidal distortions arising from relative motion between gravitationally attracting bodies will also transfer angular momentum across the system and cause the dissipation of energy. Even if these effects are small, as they can be for asteroidal bodies\cite{goldreich}, they are pervasive and will cause continual dissipation of energy for systems that are not in a relative equilibrium state. 

The ubiquity and pervasiveness of energy dissipation in the solar system and its role in the long-term evolution of bodies of all sizes motivates the main question concerning the minimum energy states for celestial mechanics systems at a given value of angular momentum. 

\section{The Sundman Inequality, Amended Potential, and Minimum Energy Function}

\subsection{The Sundman Inequality}

The fundamental result for this study is found by application of the Sundman Inequality. For generality it is applied to the integral form of the angular momentum vector, given in Eqn.\ \ref{eq:hint}. As this is not the usual form of the equations, a descriptive proof is given that relies on the Cauchy-Schwarz Inequality.

\begin{theorem}
{ $H^2 \le 2I_H T \le 2I_p T$ across all of the differential mass formulations defined in Eqn.\ \ref{eq:dmass} and system moments of inertia defined in Eqns.\ \ref{eq:ipint} and \ref{eq:ih}. }
\end{theorem}
The outermost inequality is the usual Sundman Inequality, but the sharper limits are new and are distinct for the full body problem. 
\begin{proof}
 First prove the inequality $I_H \le I_p$, which establishes the ordering on the right. First ignore the center of mass terms and only consider the moments of inertia. Taking each term independently leaves $\hat{\bfm{H}}\cdot\bfm{A}_i\cdot\bfm{I}_i\cdot\bfm{A}_i^T\cdot\hat{\bfm{H}} \le \frac{1}{2} \mbox{Trace}(\bfm{I}_i)$. If $I_1 \le I_2 \le I_3$ are the principal moments of inertia of $\bfm{I}_i$, then $\hat{\bfm{H}}\cdot\bfm{A}_i\cdot\bfm{I}_i\cdot\bfm{A}_i^T\cdot\hat{\bfm{H}} \le I_3$. Thus the inequality reduces to $I_3 \le \frac{1}{2}\left(I_1 + I_2 + I_3\right)$ or $I_3 \le I_1 + I_2$. However, this inequality and all of its permutations are a fundamental property of mass distributions and moments of inertia.
Next, consider the center of mass terms, leading to the following inequality for each term: $r_i^2 - (\hat{\bfm{H}}\cdot\bfm{r}_i)^2 \le r_i^2$, which can be trivially shown to be true. 

To finish, consider the inequality $H^2 \le 2TI_H$, which is not the usual Sundman Inequality, using the general mass integral form of the angular momentum.

First recall that $\bfm{H} = H \hat{\bfm{H}} = \int_{\cal B} \bfm{r}\times\dot{\bfm{r}} \ dm$. Dotting both sides by the (constant) unit vector aligned with the angular momentum vector yields the equality
\beq
	H & = & \int_{\cal B} \hat{\bfm{H}} \cdot (\bfm{r}\times\dot{\bfm{r}}) \ dm \\
	& = & \int_{\cal B} \dot{\bfm{r}}\cdot(\hat{\bfm{H}} \times \bfm{r}) \ dm
\eeq
Now apply the triangle inequality to the integral to find:
\beq
	\int_{\cal B} \dot{\bfm{r}}\cdot(\hat{\bfm{H}} \times \bfm{r}) \ dm & \le & \int_{\cal B} \left| \dot{\bfm{r}}\right| \left| \hat{\bfm{H}}\times\bfm{r}\right| dm 
\eeq
Squaring the original term, equal to $H^2$, and applying the Cauchy-Schwarz inequality yields the main result.
\beq
	H^2 & \le & \left[ \int_{\cal B} \left| \dot{\bfm{r}}\right| \left| \hat{\bfm{H}}\times\bfm{r}\right| dm \right]^2 \\
	\left[ \int_{\cal B} \left| \dot{\bfm{r}}\right| \left| \hat{\bfm{H}}\times\bfm{r}\right| dm \right]^2 & \le & \left[ \int_{\cal B} (\hat{\bfm{H}}\times\bfm{r})\cdot(\hat{\bfm{H}}\times\bfm{r}) \ dm \right] \left[ \int_{\cal B} \dot{{r}}^2 \ dm \right] 
\eeq
But $\int_{\cal B} \dot{{r}}^2 \ dm = 2T$ and $\int_{\cal B} (\hat{\bfm{H}}\times\bfm{r})\cdot(\hat{\bfm{H}}\times\bfm{r}) \ dm = \hat{\bfm{H}}\cdot \int_{\cal B} -\tilde{\bfm{r}}\cdot\tilde{\bfm{r}} \ dm \cdot\hat{\bfm{H}} =\hat{\bfm{H}}\cdot\bfm{I}\cdot\hat{\bfm{H}} = I_H$. Thus, $H^2 \le 2 T I_H$. 
\end{proof}

Note that for the point mass density distribution $I_p = I_H$ if all of the bodies lie in a single plane perpendicular to $\hat{\bfm{H}}$, as the moments of inertia are identically zero for a point mass. 

\subsection{The Minimum Energy Function and the Amended Potential}

The Sundman Inequality provides an important, and sharp, lower bound on the system energy for a given angular momentum. The derivation of this is simple, but the result has not been extensively used.

\begin{theorem} 
The total system energy $E = T+U$ is bounded below by the minimum energy function, defined as ${\cal E} = \frac{H^2}{2I_H} + U$, which is only a function of the system total angular momentum and the relative configuration of the components within in the system. Thus, given a total angular momentum for the system, $H$, the system energy is constrained by ${\cal E} \le E$. Equality of the system energy and the minimum energy function occurs if and only if the minimum energy function is stationary with respect to all possible variations, and corresponds with the system being in a relative equilibrium.  
\end{theorem}

\begin{proof}
In the updated Sundman Inequality, $H^2 \le 2TI_H$, replace the kinetic energy with $T = E - U$, where $E$ is the total system energy and $U$ is the gravitational potential of the system. Rearrangement of the terms yields the result $\frac{H^2}{2I_H} + U \le E$, and also defines the minimum energy function. From Eqns.\ \ref{eq:ih} and \ref{eq:udef} note that the minimum energy function is only a function of the system configuration relative to itself, and does not involve any externally defined reference points.

The relative equilibrium result arises from a theorem by Smale \cite{smaleI, smaleII}, proven in a more direct manner by Arnold in \cite{arnoldIII}. First, note that the minimum energy function is related to the ``amended potential'' (with possible differences depending on the form of the polar moment of inertia used) which has found many uses in the derivation and development of constraints on motion in the point mass $N$-body problem. In \cite{smaleII} it is proven that the stationary points of the amended potential correspond to relative equilibria of the corresponding point-mass $N$-body system at a given level of angular momentum. In \cite{simo_marsden} the proof is extended to more general dynamical systems with symmetry, which contains the Full Body system (as established in \cite{cendra_marsden})\footnote{Note that the initial, direct application of the Cauchy-Schwarz inequality yielded a polar moment of inertial $I_p$. If this moment were used to derive the minimum energy function it would no longer equal the amended potential except for the special case of a point-mass system. The modified version of the polar moment, $I_H$ is seen to precisely yield the amended potential when finite density considerations are taken into account.}. 

To finish, equality of $E$ and ${\cal E}$ is proven at a relative equilibrium. A collection of bodies in a relative equilibrium will all rotate at a constant rate and maintain a constant polar moment of inertia \cite{saari_constant_polar}. This implies that the angular momentum vector is an eigenvector of the locked inertia matrix and the value $I_H$ is an eigenvalue of this matrix (see also \cite{wang, maciejewski, scheeres_gravgrad}). Thus the total angular momentum is $\bfm{H} = \bfm{I}\cdot\bfm{\Omega} = I_H \Omega\hat{\bfm{H}}$ and the kinetic energy is $T = \frac{1}{2} \bfm{\Omega}\cdot\bfm{I}\cdot\bfm{\Omega} = \frac{1}{2} I_H \Omega^2$. Substituting into the minimum energy function yields ${\cal E} = I^2 \Omega^2 / (2I_H) + U = I \Omega^2 / 2 + U = T + U = E$. 

Now consider the case when the system is not in a relative equilibrium yet the minimum energy function equals the system energy, $\frac{H^2}{2I_H} + U = E$. Since the system is not in a relative equilibrium, the minimum energy function (amended potential) is not at a stationary value. Thus, it is possible to take an allowable variation in the configuration to increase its value, or ${\cal E} + \delta{\cal E} > {\cal E} = E$. However, then the Sundman Inequality is violated, forming a contradiction. 
\end{proof}

\begin{corollary} 
The energy is bounded by an additional function as: $\frac{H^2}{2I_p} + U \le \frac{H^2}{2I_H} + U  \le E$.
\end{corollary}

\begin{proof}
This follows immediately from the inequality $I_H \le I_p$. 
\end{proof}

\subsection{Existence of Global Energy Extrema}

Given the defined minimum energy function a rigorous approach can be formulated to the original question of what the global minimum energy configurations of a celestial mechanics system are. This, it turns out, is strongly dependent on the density distribution assumed. The following two theorems give explicit results for point mass and finite density distributions.

\begin{theorem}
For the Point Mass Density Distribution: \\
i) the minimum energy function is undefined for $N=1$, \\
ii)  the minimum energy function has a unique global minimum for $N=2$, \\
iii) the minimum energy function does not have a global minimum, or even a local minimum, for $N\ge 3$. 
\end{theorem}

\begin{proof}

{\it i}) For $N=1$, $I_p = 0$, $U = 0$ and the angular momentum is identically zero for the assumed barycentric coordinate frame. Hence the minimum energy function cannot be constructed. \\
{\it ii)} For $N=2$ a purely constructive proof is given. At $N=2$ the minimum energy function terms are
\beq
	I & = & \frac{m_1 m_2}{m_1+m_2} r^2 \\
	U & = & - \frac{{\cal G} m_1 m_2}{r}
\eeq
leading to 
\beq
	{\cal E} & = & \frac{H^2(m_1+m_2)}{2 m_1 m_2 r^2} - \frac{{\cal G} m_1 m_2}{r}
\eeq
and is defined completely by one degree of freedom, the distance between the two mass points. Figure \ref{fig:PM_Ed} shows a generic graph of this function, and it is clear that only a single extremum exists. 
\begin{figure}[ht!]
\centering
\includegraphics[scale=0.18]{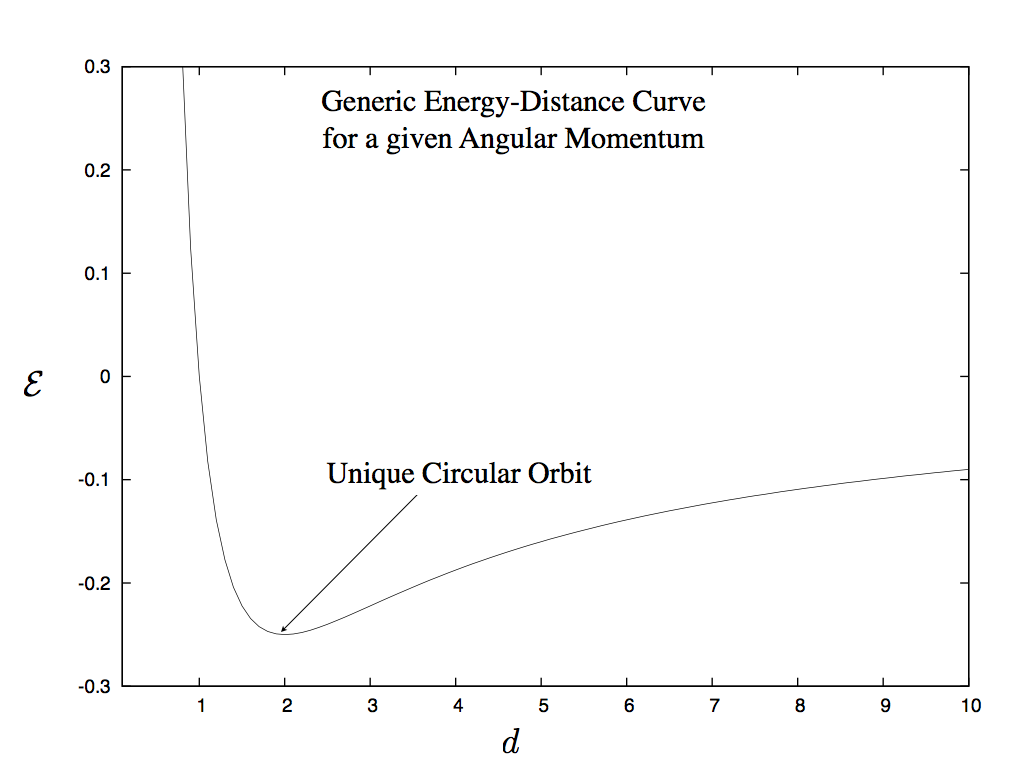}
\includegraphics[scale=0.18]{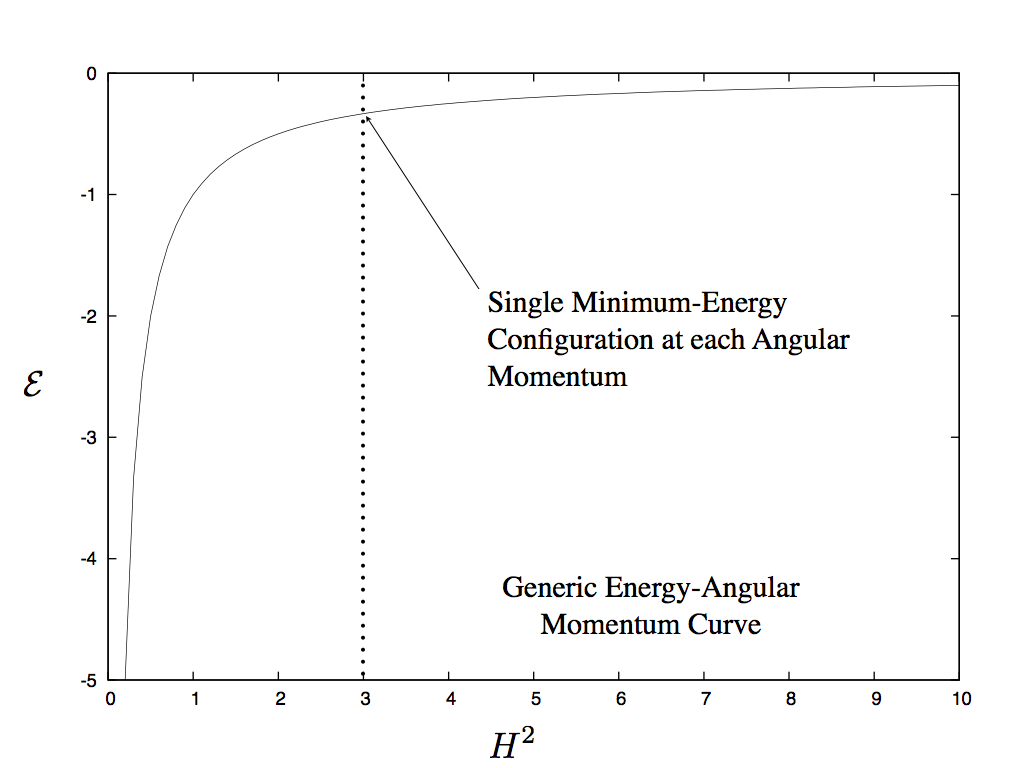}
\caption{A generic graph of the minimum energy function as a function of the distance (left) and of the angular momentum (right). For each angular momentum there is a single extrema, a minimum, which corresponds to a unique circular orbit.}
\label{fig:PM_Ed}
\end{figure}
Taking the variation of ${\cal E}$ only involves this term and yields
\beq
	\delta{\cal E} & = &  \frac{H^2(m_1+m_2)}{m_1 m_2 r^3} \left[ - 1 + \frac{{\cal G} (m_1 m_2)^2 r}{H^2(m_1+m_2)}\right] \delta r
\eeq
Setting this to zero for all $\delta r$ leads to a unique solution for $r^* \in (0,\infty)$
\beq
	r^* & = & \frac{H^2 (m_1+m_2)}{{\cal G} (m_1 m_2)^2}
\eeq
The corresponding energy is
\beq
	{\cal E}^* & = & - \frac{{\cal G}^2 (m_1 m_2)^3}{2H^2(m_1+m_2)}
\eeq
and becomes arbitrarily large as $H^2\rightarrow 0$ and goes to 0 as $H^2 \rightarrow \infty$. Figure \ref{fig:PM_Ed} shows a generic plot of this relation, note that at every angular momentum there is only a single energy configuration. 
To test for a minimum take the second variation and substitute the nominal solution to find
\beq
	\delta^2{\cal E} & = & \frac{{\cal G} (m_1 m_2)}{r^3}
\eeq
which is strictly positve, and hence the relative equilibrium is at least a local minimum of the energy function. To prove that this is a global minimum, it can be shown that ${\cal E} - {\cal E}^* \ge 0$ for all $r$. To establish this note that this inequality defines a quadratic equation in $1/r$ which can be explicitly factored to show that ${\cal E}^*$ is a global minimum at a given value of angular momentum. In terms of usual orbital mechanics this global minimum is equal to a circular orbit. 

{\it iii)}
For $N\ge 3$, note that the relative equilibrium are all stationary values of ${\cal E}$, and that ${\cal E} = E$ at the relative equilibrium. By definition these are all central configurations with a specific system rotation rate so that they remain in a relative equilibrium, and have been analyzed extensively by many authors. Moeckel provides a proof that the second variation of the energy at a fixed value of angular momentum is indefinite for all central configurations of the $N\ge3$ problem \cite{moeckel_central}. Thus, there always exist both positive and negative eigenvalues of this second variation, meaning that relative equilibrium are never even local minima of the minimum energy function, which is equal to the energy at a fixed level of angular momentum at a stationary point\footnote{This is not surprising, as if the second variation were definite it would be trivial to constructively prove stability for at least some relative equilibria -- implying that the KAM theorem would not be needed.}. This result removes any and all central configurations and related relative equilibria from consideration as minimum energy configurations. 

To completely establish that there are no global minimum it can be shown that the minimum energy function is unbounded from below for any given $H^2 \in (0,\infty)$. Again this can be done by construction. For any $N\ge3$ consider a random spatial distribution of the point masses, but allow two of them, $m_1$ and $m_2$ to be in close proximity, so that their relative distance is $r_{12} = \epsilon$. Then the minimum energy function can be split into two terms, ${\cal E} = {\cal E}' - \frac{{\cal G}m_1 m_2}{\epsilon}$, where ${\cal E}'$ is finite as long as at least one pair of bodies are not arbitrarily close to each other. Then, as the bodies are point masses let $\epsilon \rightarrow 0$ to find explicitly that ${\cal E}\rightarrow -\infty$. Thus the minimum energy function is unbounded from below for an allowable configuration and hence cannot have a global minimum across all possible system configurations. 
\end{proof}

For a {\it mechanics} system, this lack of a global minimum energy configuration is problematic, as all mechanical systems are expected to have minimum energy states. For example, this implies that for the three body problem with $H^2 > 0$ there are two final states in the presence of mechanical energy dissipation. First is that one body is ejected, leaving a two-body system and a solo body, with at least one with a well defined minimum energy state. Second is that the system remains bound and, under energy dissipation, two of the components eventually spiral towards each other until a singularity occurs. This latter system is unacceptable in a Newtonian system, as the body interactions will go beyond the validity of Newtonian gravity and physics at some point during their final spiral. 

The problem with this picture lies in the point mass density distribution traditionally used for celestial mechanics systems. Now consider minimum energy configurations for the finite density distribution.

\begin{theorem}
The minimum energy function for an $N$-body finite density distribution has a global minimum energy state for every level of angular momentum and for all $N \ge 1$. 
\end{theorem}

\begin{proof}
This theorem can be proven by showing that all of the mutual potentials $U_{ij}$ and $H^2/(2I_H)$ have global extrema across their domains. Then the sum of these functions will have global extrema across their combined domains. \\

For $N=1$ mutual gravitational potentials are undefined (as self potentials are neglected) and $U = 0$. \\

For $N\ge 2$ mutual potentials between each of the bodies in a collection are continuous functions of their domain, which is the relative position vector $\bfm{r}_{ij}$ and the relative attitude, as can be specified by the transformation dyadic $\bfm{A}_{ij}$. Let us break up the relative position vector into the unit vector and its magnitude, $\bfm{r}_{ij} = r_{ij} \hat{\bfm{r}}_{ij}$. Then the elements $\hat{\bfm{r}}_{ij}$ and $\bfm{A}_{ij}$ are defined over compact spaces, since these live in SO(2) and SO(3). For small values of $r_{ij}$ there may be additional constraints on the allowable contact locations and orientations between the bodies, but these are all continuously connected to the domain of interest (per the earlier definitions) and define closed sets. For $r_{ij}$, the lower limit of separation was defined earlier to be $d_{ij}$. Thus the domain $r_{ij} \in [d_{ij}, \infty)$, and is only closed at its lower end. Also note that all mutual potential reach a finite limit of 0 for $r_{ij} \rightarrow \infty$. Thus, a one-point compactification can be made to make the interval over which $r_{ij}$ is defined compact. Note that $U_{ij}$ is a well defined and continuous function across this domain. Then by the extreme value theorem $U_{ij}$ has a global minimum and maximum. 

Now consider the function $H^2 / (2 I_H)$. The moment of inertia is always $> 0 $ for a finite density distribution, as even when $N=1$ it equals the moment of inertia about the direction $\hat{\bfm{H}}$, always non-zero by definition. As any $r_{ij}\rightarrow\infty$ the function $1/I_H \rightarrow 0$, and thus is well defined. As before, one can argue that $1/I_H$ is continuous over its interval of definition and that its domain is compact (or can be made compact). Thus, by the extreme value theorem $H^2 / (2I_H)$ has a global minimum and maximum.

To complete the proof, trivially note that the addition of continuous functions defined over a common compact domain remains continuous. Thus the extreme value theorem applies to the minimum energy function ${\cal E} = \frac{H^2}{2I_H} + \sum_{i=1}^{N-1}\sum_{j=i+1}^N U_{ij}$, and it has a global minimum and maximum. 
\end{proof}

\section{Minimum Energy Configurations in the \\
Spherical Full Body Problem}

Having proven existence for the main result, its implications can be explored using the simplest possible extension of the point-mass celestial mechanics problem to the finite density case, which is the spherical full body problem. See previous works by \cite{wang, maciejewski, scheeres_F2BP, scheeres_gravgrad, fahnestock_icarus, scheeres_F2BP_planar} for more detailed Celestial Mechanics discussions of the Full 2-Body problem for non-spherical bodies. The following analysis strives to be more complete by analyzing all possible options, enabled in part by the simpler model of interaction that the spherical restriction allows. 

\subsection{$N=1$}

The mutual potential is identically zero and the moment of inertia of the sphere is equal through any axis through its center. Thus the minimum energy function takes on the simple form
\beq
	{\cal E} & = & \frac{H^2}{m d^2/5}
\eeq
and has no degrees of freedom. Thus this system is always in a minimum energy state. Had the sphere restriction not been applied, then the minimum energy state would have the body reoriented so that the maximum moment of inertia would lie along $\hat{\bfm{H}}$ \cite{burns_safronov}.

\subsection{$N=2$}

Now revisit the two-body problem for the case of finite density, focusing on relative equilibrium, local minimum configurations and globally minimum configurations. This problem shows that extensions to finite density mass distributions fundamentally changes the energy structure of this problem and yields the non-intuitive result that there can be multiple circular orbits at a given angular momentum and that there can be unstable circular orbits. 
For generality in the following the two bodies have different masses, $m_1$ and $m_2$, and diameters, $d_1$ and $d_2$.

\begin{theorem}
\label{thm:N2orb}
There are 0, 1 or 2 orbital relative equilibria for the 2-body problem as a function of angular momentum. When there are 2 orbital equilibria, one is energetically stable and the other is energetically unstable.
\end{theorem}

\begin{proof}
The potential and polar moments of inertia are:
\beq
	U & = & -\frac{{\cal G}m_1 m_2}{r} \\
	I_H & = & \frac{m_1 m_2}{m_1 + m_2} r^2 + I_s \\
	I_s & = & \frac{1}{10} \left( m_1 d_1^2 + m_2 d_2^2\right) \\
	r & \ge & \frac{d_1 + d_2}{2} = d_{12}
\eeq
Normalize this system by 
\beq
	\overline{r} & = & r / d_{12} \\
	\overline{H} & = & \frac{H\sqrt{ m_1 + m_2}}{\sqrt{{\cal G} d_{12}} m_1 m_2}  \\
	\overline{I}_s & = & \frac{m_1 + m_2}{ d_{12}^{2} m_1 m_2} I_s \\
	\overline{U} & = & \frac{U d_{12}}{{\cal G} m_1 m_2}
\eeq
In the following the $\overline{(-)}$ notation will be dropped with the assumption that all quantities are normalized as given above.
With this normalization the range of allowable distances between the particles is defined by $r \ge 1$. 

The minimum energy function is then
\beq
	{\cal E} & = & \frac{H^2}{2\left( r^2 + I_s\right)}  - \frac{1}{r} 
\eeq
This is a single degree of freedom function so relative equilibria are found by taking a variation with respect to $r$
\beq
	\delta{\cal E} & = & \left[ - \frac{H^2}{\left( r^2 + I_s\right)^2}  + \frac{1}{r^3}\right] r \delta{r}
\eeq
Setting the variation equal to zero yields an equation for $r$:
\beq
	r^4 - {H}^2 r^3 + 2 {I}_s r^2 + {I}_s^2 & = & 0 \\
\eeq

Applying the Descartes rule of signs the polynomial either has 2 or no solutions, a drastic change from the point-mass problem which always had a single solution\footnote{If ${I}_s \rightarrow 0$ the single solution from the point-mass case is recovered.}. Of particular interest is the point where the transition from zero to 2 solutions occurs, which will be a double root of this polynomial. Set the derivative to zero to find
\beq
	\left( 4 r^2 - 3 {H}^2 r + 4 {I}_s \right) r& = & 0 
\eeq
Evaluating both equations simultaneously shows that the bifurcation from zero to two solutions occurs at
\beq
	r^* & = & \sqrt{3{I}_s} \label{eq:rstar} \\
	{H}^2 & = & \frac{16}{3\sqrt{3}} \sqrt{{I}_s} \label{eq:Hstar}
\eeq
Note that the minimum value of ${I}_s$ occurs when $m_1 = m_2$ which leads to $d_1 = d_2$ under an equal density assumption, which leads to ${I}_s \ge 0.4$. Thus, $r^* \ge \sqrt{1.2} > 1$ for all mass values with constant density and the bifurcation structure for $N=2$ is robust and always occurs when the bodies are separate from each other. 

An alternative approach is to solve for ${H}^2$ as a function of $r$, yielding
\beq
	{H}^2 & = & \frac{(r^2 + {I}_s)^2}{r^3} \label{eq:N2HofR}
\eeq
This has a minimum point, corresponding to the bifurcation of the roots to the equation given above in Eqns.\ \ref{eq:rstar} and \ref{eq:Hstar}. 
For higher values of $H$ the system will then have two roots, one to the right of $r^*$ (the outer solution) and the other to the left (the inner solution). At the bifurcation point the system technically only has one solution.

The stability of each of these solutions is determined by inspecting the second variation of ${\cal E}$ with respect to $r$, evaluated at the relative equilibrium. Taking the variation and making the substitution for $H^2$ from Eqn.\ \ref{eq:N2HofR} yields
\beq
	\delta^2{\cal E} & = & \left[ \frac{4}{(r^2+{I}_s)} - \frac{3}{r^2} \right] \frac{(\delta r)^2}{r}
\eeq
Checking for when $\delta^2{\cal E} > 0$ yields the condition $r^2 > 3{I}_s$. Thus, the outer solution will always be stable while the inner solution will always be unstable, with the relative equilibria occurring at a minimum and maximum of the minimum energy function, respectively.
\begin{figure}[ht!]
\centering
\includegraphics[scale=0.28]{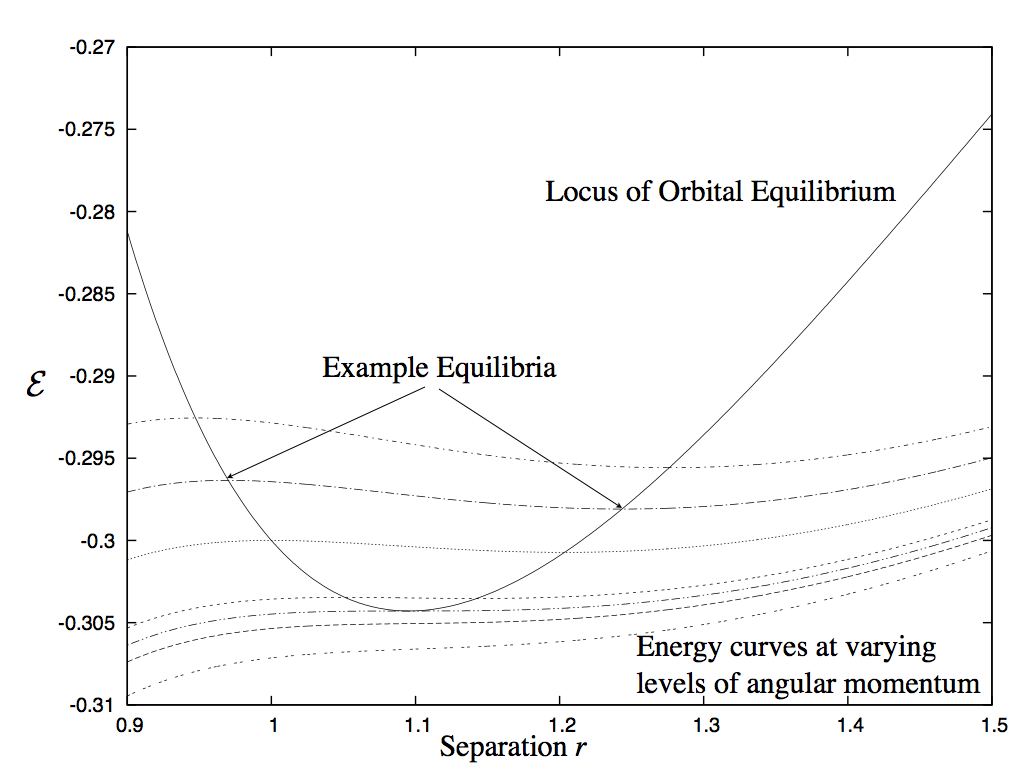}
\caption{Locus of orbital equilibria across a range of angular momentum values. Plotted is the minimum energy function versus the system configuration, $r$, the distance between the two particles. This plot assumes equal masses and sizes of the two particles. For clarity, equilibrium solutions are shown below the physical limit $r \ge 1$.}
\label{fig:N2locus}
\end{figure}
Figure \ref{fig:N2locus} shows characteristic energy curves and the locus of equilibria for different levels of angular momentum for the spherical full 2-body problem.  This should be contrasted with 
the energy function for the point mass 2-body problem, shown in Fig.\ \ref{fig:PM_Ed}, which only has one relative equilibrium. 

\end{proof}

\begin{theorem}
\label{thm:N2rest}
There are 0 or 1 unique resting relative equilibrium configurations as a function of angular momentum. When it exists, the resting equilibrium is energetically stable.
\end{theorem}

\begin{proof}
There is only a single unique contact configuration for the $N=2$ case, with the two bodies touching with their centers of mass a distance $d_{12}$ apart. In normalized coordinates the minimum energy function is ${\cal E} = \frac{{H}^2}{2(r^2 + {I}_s)} - \frac{1}{r}$ where $r=1$ signifies contact between the bodies. For a contact structure to be a resting equilibrium, any variations in its degrees of freedom must conform to the rigid body constraint (i.e., $\delta r \ge 0$ in this case) and must only cause an increase in the minimum energy function, $\delta{\cal E} \ge 0$. If a decrease in energy occurs, the system will trend away from the configuration along an allowable variation in the degree of freedom. Taking the variation in the minimum energy function at the contact configuration of $r=1$ yields
\beq
	\delta{\cal E} & = & \left[ 1 - \frac{{H}^2}{(1 + {I}_s)^2}\right] \delta r
\eeq
Since $\delta r \ge 0$ is the only variation allowed, the relative equilibrium will exist when $1 - \frac{{H}^2}{(1 + {I}_s)^2} \ge 0$. Thus if ${H} \le 1 + {I}_s$ the static resting equilibrium will exist and if ${H} > 1 + {I}_s$ there will be no resting equilibrium.

When the resting equilibrium exists, any variation in the degree of freedom will only increase the energy, meaning that its variation is positive definite and this relative equilibrium is energetically stable.
\end{proof}

\begin{figure}[ht!]
\centering
\includegraphics[scale=0.18]{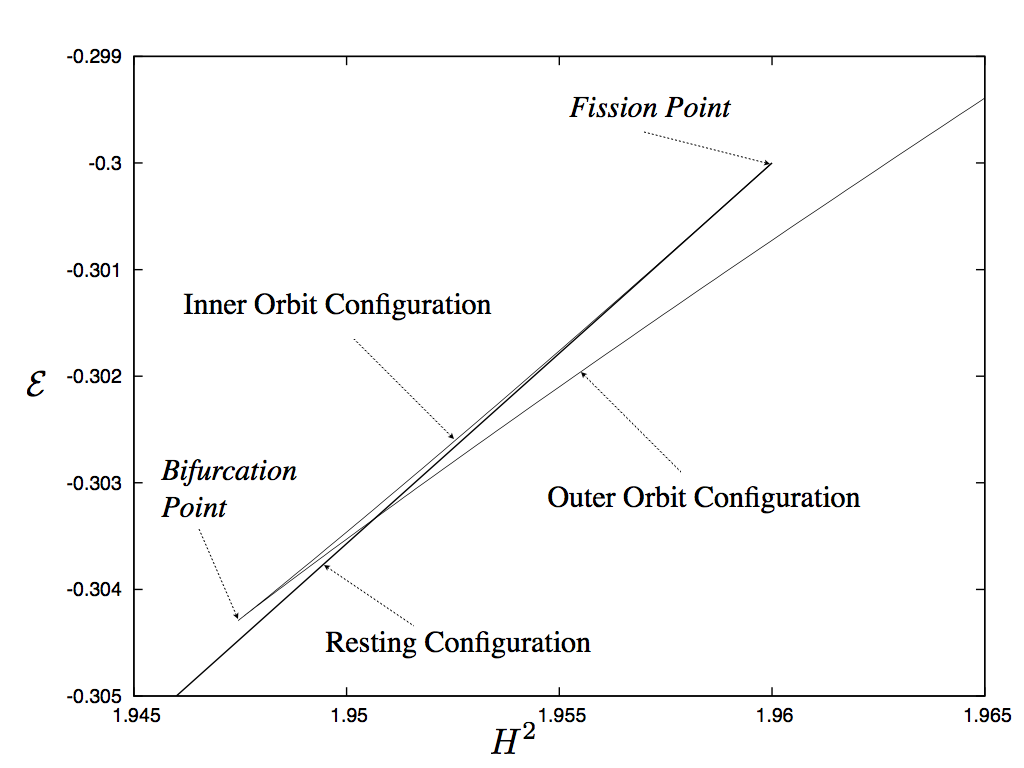}
\includegraphics[scale=0.18]{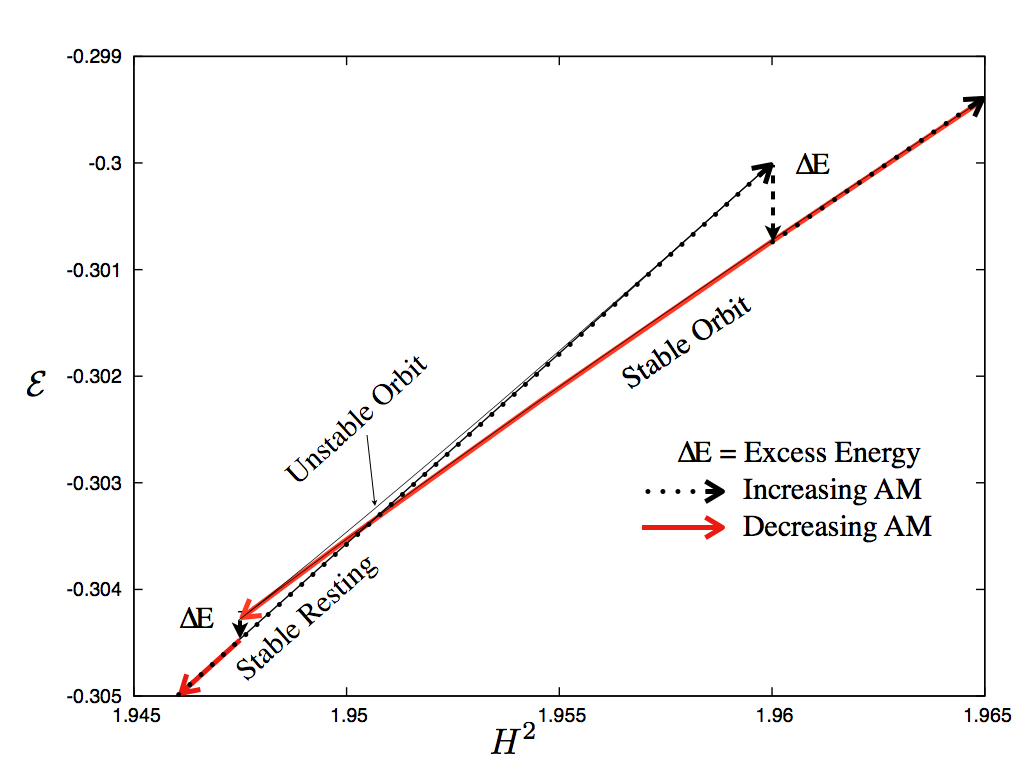}
\caption{The energy-angular momentum diagram of relative equilibria (left) and the transition diagram for increasing and decreasing angular momentum (right) in the Full 2-body problem for equal sized bodies.}
\label{fig:N2bifurcation}
\end{figure}

\begin{theorem}
The complete bifurcation chart of relative equilibria, minimum energy states, and global minimum energy states of the sphere restricted $N=2$ full body problem as a function of angular momentum follows. The sequence is graphically represented in Figure \ref{fig:N2bifurcation}. 
\begin{description}

\item
[$0 \le {H}^2 < \frac{16}{3\sqrt{3}} \sqrt{{I}_s}$] The minimum energy configuration, and the only relative equilibrium, is for the two bodies to be resting on each other with $r = d_{12}$.

\item
[$\frac{16}{3\sqrt{3}} \sqrt{{I}_s} \le {H}^2 < \frac{4\sqrt{{I}_s}(1+{I}_s)}{\sqrt{1+{I}_s} + \sqrt{{I}_s}}$] Two orbital relative equilibria bifurcate into existence at a radius $r^* = \sqrt{3{I}_s}$ and grow to larger and smaller values of $r$. The outer solution is at a local minimum of the energy, and thus is stable. The inner solution is at a local maximum of the energy, and thus is unstable. The global minimum energy configuration remains the contact solution.

\item
[$\frac{4\sqrt{{I}_s}(1+{I}_s)}{\sqrt{1+{I}_s} + \sqrt{{I}_s}} \le {H}^2 < (1+{I}_s)^2$] The situation is the same as above, except now the outer orbital solution becomes the global minimum, and the resting solution becomes a local minimum.

\item
[$(1+{I}_s)^2 = {H}^2$] The inner orbital solution collides with the resting solution and they annihilate each other. This can be called the ``fission'' condition for the resting configuration, as it is the point beyond which the resting condition no longer exists and the particles that were resting on each other are forced to enter orbit about each other. 

\item
[$(1+{I}_s)^2 < {H}^2$] Beyond the fission limit there is only one relative equilibrium, the outer orbital solution, which is the minimum energy configuration.

\end{description}

\end{theorem}

\begin{proof}
Theorems \ref{thm:N2orb} and \ref{thm:N2rest} outline the values of ${H}^2$ at which the relative equilibria exist and what their stability properties are. The main items to prove in the above are {\it i}) the transitions between the global minimum configuration and {\it ii}) the collision of the unstable orbital equilibrium with the resting equilibrium.

{\it i}) To prove this find the value of ${H}^2$ that gives equality of the energy of the resting and orbital configurations and verify the inequalities. 
To understand the global minimum transitions compare the energy of the resting and the stable outer orbital configurations with each other. Equating the orbital and the resting minimum energy functions at the same level of angular momentum, it is possible to solve for the angular momentum at which the two functions are equal:
\beq
	{H}^2 & = & \frac{2(1+{I}_s)(r^2+{I}_s)}{r(r+1)}
\eeq
To identify the transition point between global minima substitute for the angular momentum as a function of orbit radius (Eqn.\ \ref{eq:N2HofR}) and simplify the resulting equation to find
\beq
	(r-1)\left( r^2 - 2{I}_s r - {I}_s \right) & = & 0 \label{eq:N2eqcond}
\eeq
with an obvious root at $r=1$ (discussed in a moment) and a positive real root at $r_s = {I}_s + \sqrt{{I}_s ({I}_s +1)}$. When $r = r_s$ the global minimum will switch from the resting configuration to the outer orbital configuration as angular momentum increases. Substituting into Eqn.\ \ref{eq:N2HofR} and simplifying yields
\beq
	{H}^2_s & = & \frac{4\sqrt{{I}_s}(1+{I}_s)}{\sqrt{1+{I}_s} + \sqrt{{I}_s}} 
\eeq

Next check the inequalities $\frac{16}{3\sqrt{3}} \sqrt{{I}_s} < \frac{4\sqrt{{I}_s}(1+{I}_s)}{\sqrt{1+{I}_s} + \sqrt{{I}_s}} < (1+{I}_s)^2$. First note that if ${I}_s$ becomes large the inequalities devolve into $\frac{16}{3\sqrt{3}} \sqrt{{I}_s} < 2 {I}_s < {I}_s^2$ and strictly hold for ${I}_s > 64/27$ for the lower inequality and ${I}_s > 2$ for the upper. Also, taking the minimum value of ${I}_s = 0.4$ and inserting it into the relationships yields $1.9474\ldots < 1.9506\ldots < 1.96$. To verify over the remaining interval it suffices to plot and compare the functions over this compact set.

{\it ii}) Note that the fission of the resting configuration will occur when the inner, unstable relative equilibrium touches the value $r = 1$. From Eqn.\ \ref{eq:N2eqcond} note that the energy of the resting and orbital relative equilibria are equal at this point. Evaluating the angular momentum at this point then yields the value ${H}^2 = (1+{I}_s)^2$.
\end{proof}

\begin{corollary}
The $N=2$ transition diagram as a function of angular momentum is discontinuous and exhibits hysteresis. To track the evolution of a system under changing angular momentum will require dynamical analysis at specific transition points.
\end{corollary}

\begin{proof}
In Figure \ref{fig:N2bifurcation} the energy of the relative equilibria is graphed when they are stable as a function of angular momentum for equal-sized bodies (with qualitatively similar results holding across all values). Under increasing or decreasing angular momentum the configuration will remain in a locally stable relative equilibrium after the global minimum configuration has switched to a different configuration. Thus, when the local relative equilibrium becomes unstable or ceases to exist, there is a finite energy difference between the current system state and its minimum energy state. If angular momentum is increased beyond this point this directly implies a period of dynamical interaction during which energy dissipation may occur, eventually leading to the minimum energy state. 

At fission the system has a corresponding energy of $\frac{1}{2}( {I}_s - 1)$ and thus at fission the energy can be positive if ${I}_s > 1$. For constant density spheres this will occur when the ratio between the masses is $<  0.204\ldots$ or $> 4.902\ldots$, assuming constant density \cite{scheeres_F2BP_planar}. In \cite{scheeres_F2BP} it is proven that full body systems can escape if their free energy (equal to the energy as defined in this paper) is positive. Thus, across the range of different mass ratios it is even possible for the resulting fission system to eventually escape prior to sufficient energy dissipation for being trapped into the outer orbital configuration. In \cite{pravec_fission} this same mass ratio is found in a class of natural bodies that are expected to have undergone fission in the past.
\end{proof}


For the Spherical Full 2-body problem, these diagrams are notional as the actual dynamics of these systems consist of purely elliptic motions. Indeed, without explicit energy dissipation the excess energy can never be depleted and the system will not evolve. Also, the rotational motion of the spheres are completely decoupled from orbital motion in the rigid body limit. However, these diagrams present the underlying energy structure for fixed levels of angular momentum. Even the slightest departure from rigidity will enable energy dissipation to occur if perturbed from either of the relative orbital equilibria. If perturbed from the stable equilibrium the dissipation will allow the system to settle into its minimum energy state. If perturbed from the unstable equilibrium it will either evolve towards the contact relative equilibrium, the stable orbital one or may escape before it reaches this limit if the total energy is positive. 

\subsection{$N=3$}

This discussion is restricted to bodies having equal sizes and densities. Thus, all particles have a common spherical diameter $d$ and mass $m$. A restriction to the planar problem is also made to simplify the analysis, although non-planar configurations can also be analyzed.
For the equal mass, point-mass celestial mechanics problem, there are only two known relative equilibrium configurations, the Euler and Lagrange solutions. When finite densities are incorporated there are at least 6 orbital configurations and a total of 9 distinct relative equilibria. If non-equal size bodies of similar density are considered, there will be additional distinguishable relative equilibria, but this analysis is left for future work. 

Given this restriction the polar moment and potential energy take on simpler forms. 
\beq
	I_H & = & \frac{m}{3} \left( r_{12}^2 + r_{23}^2 + r_{31}^2 \right) + \frac{3}{10} m d^2 \\
	U & = & - {{\cal G}m^2} \left[ \frac{1}{r_{12}} + \frac{1}{r_{23}} + \frac{1}{r_{31}} \right]
\eeq
where $m$ is the common mass of each body and $d$ the common diameter. Then $r_{ij} \ge d$ for all of the relative distances.  Now introduce some convenient normalizations, scaling the moment of inertia by $m d^2$ and scaling the potential energy by ${\cal G}m^2 / d$. Then the minimum energy function is
\beq
	\overline{\cal E} & = & \frac{\overline{H}^2}{2 \left[ \left(\overline{r}_{12}^2 + \overline{r}_{23}^2 + \overline{r}_{31}^2\right)/3 + 0.3 \right]} 
		- \left[ \frac{1}{\overline{r}_{12}} + \frac{1}{\overline{r}_{23}} + \frac{1}{\overline{r}_{31}} \right] 
\eeq
where
\beq
	\overline{\cal E} & = & \frac{{\cal E} d}{{\cal G}m^2} \\
	\overline{H}^2 & = & \frac{H^2}{{\cal G}m^3 d}
\eeq
and the constraint from the finite density assumption becomes $\overline{r}_{ij} \ge 1$.

In the following the $\overline{-}$ notation is dropped for $r_{ij}$ and $H$, as it will be assumed that all quantities are normalized. 

\begin{figure}[ht!]
\centering
\includegraphics[scale=0.28]{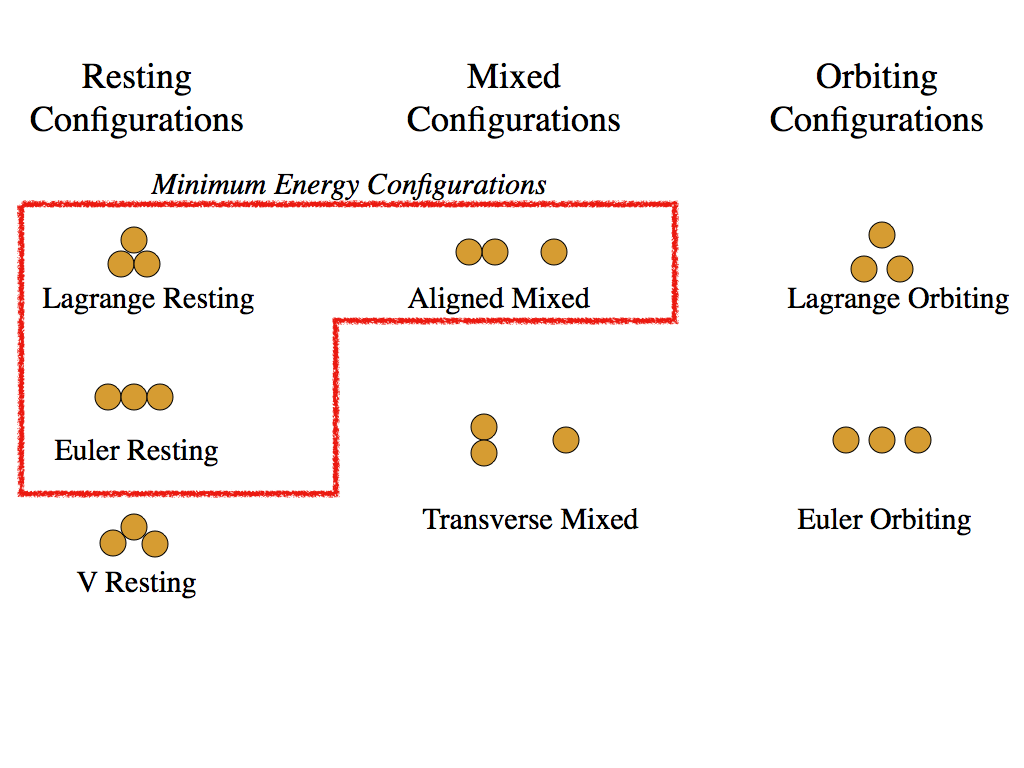}
\caption{Diagram of all relative configurations for the Full 3-body problem. Those boxed can also be stable, and each takes a turn as the global minimum energy configuration for a range of angular momentum.}
\label{fig:N3diagram}
\end{figure}

\begin{theorem}
There are a total of 9 unique (under symmetry transformations) relative equilibria that can exist in the $N=3$ sphere restricted full body problem. These are shown in Figure \ref{fig:N3diagram} and described below. 

\begin{enumerate}
 
\item
There are two static resting relative equilibrium configurations, the Euler Resting and Lagrange Resting configurations. The Lagrange Resting configuration is energetically stable whenever it exists while the Euler Resting configuration transitions from unstable to stable as a function of angular momentum while it exists.

\item
There is one dynamic resting relative equilibrium configurations, the ``V'' configuration, which is always energetically unstable when it exists.

\item
There are 4 mixed orbital and resting relative equilibrium configurations, the inner and outer aligned and transverse mixed configurations. The 2 inner configurations and the outer transverse configurations are always energetically unstable. The outer aligned configuration is always stable. 

\item
There are 2 purely orbital relative equilibrium configurations, the Lagrange and Euler configurations. These are always energetically unstable.
\end{enumerate}

\end{theorem}

\begin{proof}
\begin{enumerate}
\item
{\bf Static resting configurations}

First consider the static resting configurations, defined as when all of the bodies are in contact and maintain a fixed shape over a range of angular momentum values. When in contact there is only one degree of relative freedom for the system, defined as the angle between the two outer particles as measured relative to the center of the middle particle and shown in Fig.\ \ref{fig:contact_geometry}.
As defined the angle must always lie in the limit $60^\circ \le \theta \le 300^\circ$. 

\begin{figure}[ht!]
\centering
\includegraphics[scale=0.18]{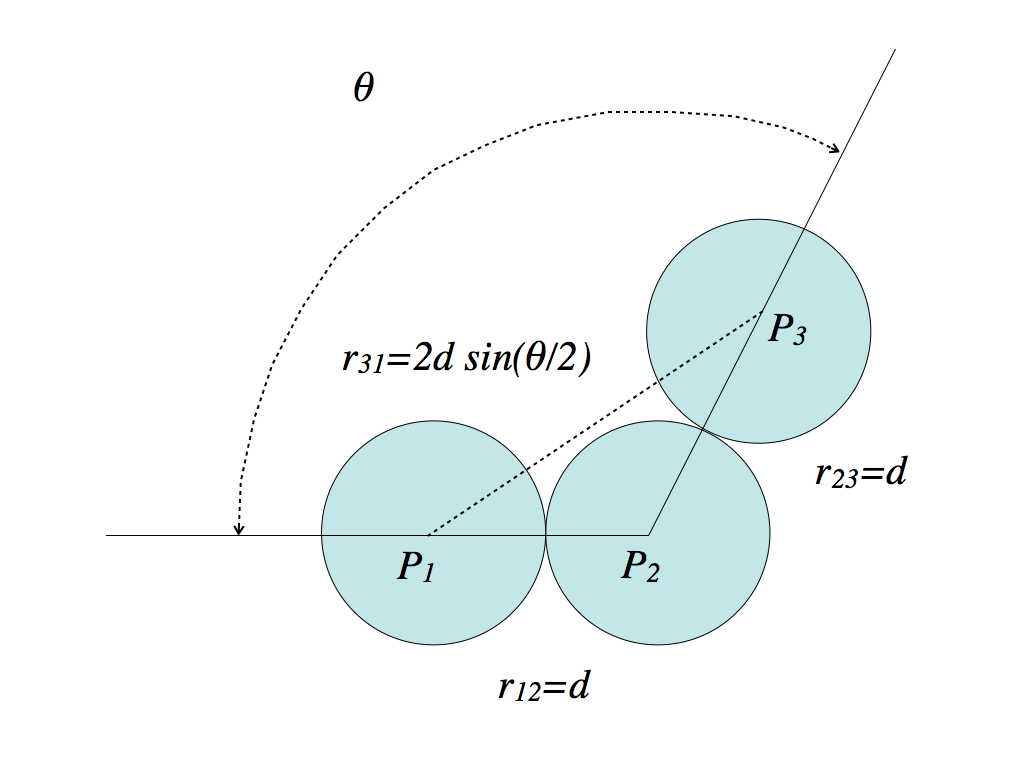}
\includegraphics[scale=0.18]{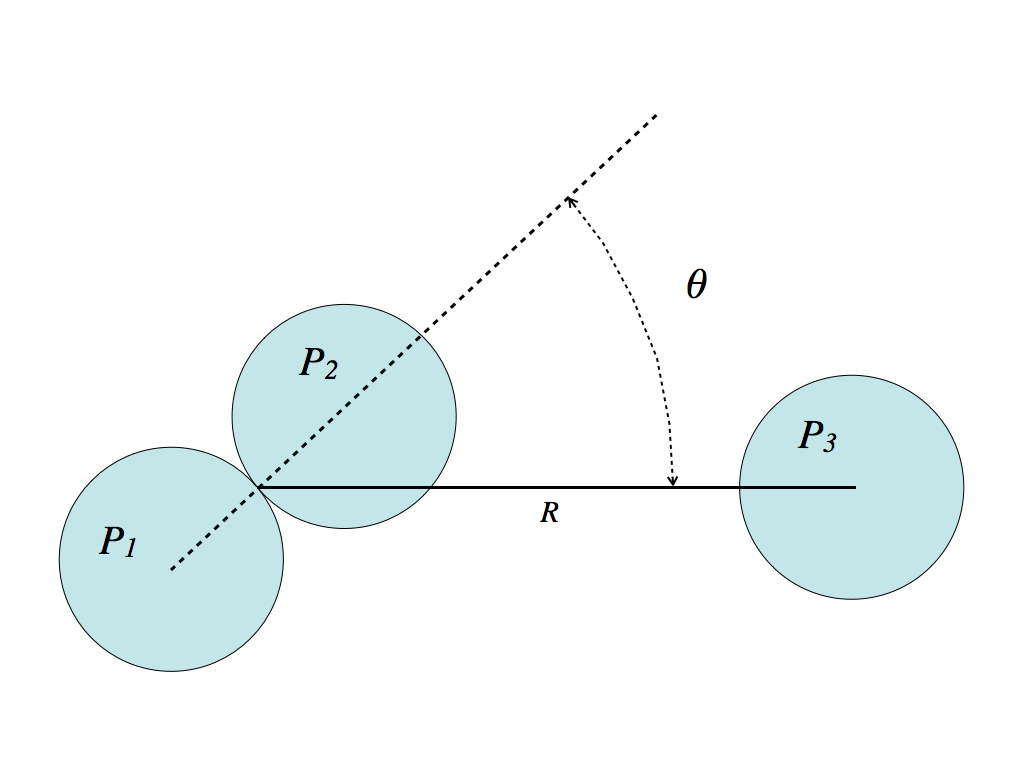}
\caption{Generic description of the planar contact geometry between three equal sized particles (left) and the mixed configuration geometry (right).}
\label{fig:contact_geometry}
\end{figure}

Given the geometric relationships in Fig.\ \ref{fig:contact_geometry}, the minimum energy function can be written as
\beq
	{\cal E}_S & = & \frac{H^2}{2\left[ \left( 2 + 4\sin^2(\theta/2)\right)/3 + 0.3\right]} - 2 - \frac{1}{2\sin(\theta/2)} 
\eeq
where the $S$ subscript stands for ``Static.''
The first variation is then
\beq
	\delta_{}{\cal E}_S & = & \sin\theta \left[ - \frac{3 H^2}{\left(2.9+4\sin^2(\theta/2)\right)^2} + \frac{1}{8\sin^3(\theta/2)} \right] \delta\theta \label{eq:rest_cond}
\eeq
The second variations will be considered on a case-by-case basis. Since this system has a constraint on the angle $\theta$, both the free variations of $\theta$ and the constrained variations when at the limit must be considered. 

\paragraph{Euler Rest Configuration}

If $\theta = 0$ the minimum energy function will be stationary. Define this as the Euler Rest Configuration, which consists of all three particles lying in a single line. The stability of this configuration is evaluated by taking the second variation of the energy function and evaluating it at $\theta = 180^\circ$, yielding
\beq
	\delta^2{\cal E}_{S} & = & - \frac{1}{8} \left[ 1 - \frac{24}{(6.9)^2} H^2 \right] \delta\theta^2
\eeq
Recall that the stability condition is that the second variation of the energy be positive definite, yielding an explicit condition for stability as $H^2 > 1.98375$, with lower values of $H^2$ being definitely unstable. 
The Euler Rest Configuration energy can be specified as a function of angular momentum:
\beq
	{\cal E}_{SE} & = & \frac{H^2}{2\left(2.3\right)} - \frac{5}{2}
\eeq

\paragraph{Lagrange Rest Configuration}

Now consider the constrained stationary point with $\theta = 60^\circ$ ($300^\circ$). Define this as the Lagrange Rest Configuration. Here it suffices to evaluate the first variation at the boundary condition, yielding
\beq
	\left. \delta{\cal E}_{S}\right|_{60^\circ} & = & \frac{\sqrt{3}}{2} \left[ - \frac{3H^2}{(3.9)^2} + 1 \right] \delta\theta
\eeq
At the $60^\circ$ constraint $\delta\theta\ge 0$ and the Lagrange Rest Configuration will exist and be stable for $H^2 < 5.07$, but beyond this limit an increase in $\theta$ will lead to a decrease in energy and the relative equilibrium will no longer exist. Note that if the $\theta = 300^\circ$ limit is taken, the sign of the first variation switches but the constraint surface is now $\delta\theta \le 0$, and the same results hold. 

The Lagrange Rest Configuration energy can be specified as a function of angular momentum:
\beq
	{\cal E}_{SL} & = & \frac{H^2}{2\left(1.3\right)} - 3
\eeq
Comparing the energy of these two rest configurations shows that the Lagrange configuration has lower energy for $H^2 < 2.99$ while the Euler configuration has a lower energy above this level of angular momentum. 

\item
{\bf Variable Contact Configurations}

In addition to the static resting configurations it is also possible to have full contact configurations which change as the angular momentum varies. These are not fully static, as they depend on having a specific level of angular momentum, generating centripetal accelerations that balance the gravitational and contact forces. For these configurations, as the level of angular momentum varies the configuration itself shifts, adjusting to the new environment. For the $N=3$ case there is only one such ``variable contact'' configuration when restricted to the plane. This particular configuration is always unstable, yet plays an important role in mediating the stability of the other configurations.

\paragraph{``V'' Rest Configuration} 

The variable contact configuration that can exist for this system, called the ``V'' Configuration for obvious reasons, yields the final way for a stationary value of the minimum energy function to exist, with the terms within the parenthesis of Eqn.\ \ref{eq:rest_cond} equaling zero. Instead of solving the resulting quartic equation in $\sin(\theta/2)$ it is simpler to evaluate the angular momentum as a function of the system configuration to find
\beq
	H^2 & = & \frac{\left(2.9 + 4\sin^2(\theta/2)\right)^2}{24\sin^3(\theta/2)}
\eeq
The range of angular momenta that correspond to this configuration can be traced out by following the degree of freedom $\theta$ over its range of definition.
Thus the V Rest Configuration will exist for angular momentum values ranging from $H^2 = 1.98375$ at $\theta = 180^\circ$ to $H^2 = 5.07$ at $\theta = 60^\circ$. Note that the angles progress from $\theta = 180^\circ \rightarrow 60^\circ$ as the angular momentum increases, and that the limiting values occur when the Euler Rest Configuration stabilizes and the Lagrange Rest Configuration destabilizes. Note that a symmetric family moves from $\theta = 180^\circ$ to $\theta=300^\circ$ at the same levels of angular momentum.

Taking the second variation and evaluating the sign of $\delta^2{\cal E}_{C}$ along the V configuration for arbitrary variations shows that it is always negative definite over the allowable values of $\theta$ and thus that the V Rest Configuration is always unstable. 

\item
{\bf Mixed Configurations}

Now consider mixed configurations where both resting and orbital states can co-exist. For $N=3$ there is only one fundamental topology of this class allowed, two particles rest on each other and the third orbits. Further, from simple symmetry arguments two candidate states for relative equilibrium can be identified, a Transverse Configuration where the line joining the two resting particles is orthogonal to the third particle ($\theta = \pm90^\circ$), and an Aligned Configuration where a single line joins all of the mass centers ($\theta = 0, 180^\circ$). To enable a stability analysis a full configuration description of these systems is introduced which requires two coordinates: the distance from the center of the resting pair to the center of the third particle to be $R$, and the angle between the line $R$ and the line joining the resting pair as $\theta$  (see Fig.\ \ref{fig:contact_geometry}). 


The distances between the different components can be worked out as
\beq
	r_{12} & = & 1 \\
	r_{23} & = & \sqrt{ R^2 - R \cos\theta + 0.25} \\
	r_{31} & = & \sqrt{ R^2 + R \cos\theta + 0.25} 
\eeq
Thus the minimum energy function takes on the form
\beq
	{\cal E}_M & = & \frac{3H^2}{4\left(1.2 + R^2\right)} - 1 \nonumber \\
	& & - \frac{1}{\sqrt{ R^2 - R \cos\theta + 0.25}} - \frac{1}{\sqrt{ R^2 + R \cos\theta + 0.25}} 
\eeq
where the $M$ stands for ``Mixed.''
Taking the variation with respect to $\theta$ yields
\beq
	\delta_\theta{\cal E}_M & = & \frac{R\sin\theta}{2}  \times \\
	& & \nonumber \left[ \frac{1}{\left( R^2 - R \cos\theta + 0.25\right)^{3/2}} - \frac{1}{\left( R^2 + R \cos\theta + 0.25\right)^{3/2}}\right] \delta\theta
	\label{eq:delta_theta_M}
\eeq
As expected, the variation is stationary for the Aligned Configuration, $\theta = 0, 180^\circ$, and for the Transverse Configuration, $\theta = \pm90^\circ$.
The variation in the distance yields
\beq
	\delta_R{\cal E}_M & = & \left[ - \frac{3H^2 R}{2(1.2+R^2)^2}  + \frac{2R-\cos\theta}{2\left( R^2 - R\cos\theta + 0.25\right)^{3/2}} \right. \nonumber \\
	& & \left. + \frac{2R+\cos\theta}{2\left( R^2 + R\cos\theta + 0.25\right)^{3/2}}\right] 
		\delta R \label{eq:delta_R_M}
\eeq
and is discussed in the following.

\paragraph{Transverse Configurations}

First consider the Transverse Configurations with $\theta = \pm90^\circ$. Evaluating the variation of ${\cal E}_M$ with respect to $R$, setting this to zero, and substituting $\theta = 90^\circ$ allows us to solve for the angular momentum explicitly as a function of the separation distance
\beq
	H_{MT}^2 & = & \frac{4(1.2+R^2)^2}{3\left(R^2 + 0.25\right)^{3/2}} 
\eeq
This function has a minimum value of angular momentum of $H_{MT}^2 \sim 4.002\ldots$ which occurs at $R = \sqrt{2.6}$. This is an allowable value of separation and thus this bifurcation will indeed occur. For higher values of angular momentum there are two relative equilibria, one with separation less than $\sqrt{2.6}$ and the other with separation larger than this. The inner solution touches the other two particles, forming a Lagrange-like configuration, when $R = \sqrt{3}/2$. Substituting this into the above equation for $H_{MT}^2$ shows that this occurs at a value of $5.07$, which is precisely the value at which the Lagrange Rest Configuration becomes unstable. Recall that this was also the value of angular momentum at which point the V Rest Configuration terminated by reaching $60^\circ$. Thus at this value, which is also equal to the Lagrange Orbit Configuration angular momentum at this distance, the inner Transverse Configuration family of solutions terminates. Conversely, the outer Transverse Configuration persists for all angular momentum values above the bifurcation level. 

Now consider the energetic stability of this class of relative equilibria. First note that the cross partials, $\delta^2_{\theta R}{\cal E}_M$ are identically equal to zero for the Transverse Configuration. This can be easily seen by taking partials of Eqn.\ \ref{eq:delta_R_M} with respect to $\theta$ and inserting the nominal value $\theta = \pm 90^\circ$. Next, taking the second partial of Eqn.\ \ref{eq:delta_theta_M} with respect to $\theta$ and evaluating it at the nominal configuration yields 
\beq
	\delta^2_{\theta\theta}{\cal E}_{MT} & = & \frac{-3 R^2}{2\left( R^2 + 0.25\right)^{5/2}}  (\delta\theta)^2
\eeq
and $\delta^2_{\theta\theta}{\cal E}_{MT} < 0$. It is not necessary to check further as this tells us that none of the Transverse Configurations are energetically stable.
The explicit energy of the Transverse Configurations is
\beq
	{\cal E}_{MT} & = & \frac{3H^2}{4\left(1.2 + R^2\right)} - 1 - \frac{2}{\sqrt{ R^2 + 0.25}} 
\eeq

\paragraph{Aligned Configurations}

Now consider the Aligned Configurations with $\theta = 0,180^\circ$. Again solve for the angular momentum as a function of separation
\beq
	H^2_{MA} & = & \frac{2(1.2+R^2)^2}{3R} \left[ \frac{1}{\left(R-\frac{1}{2}\right)^2} + \frac{1}{\left(R+\frac{1}{2}\right)^2} \right] 
\eeq
Finding the minimum point of this equation as a function of $R$ yields a cubic equation in $R^2$ without a simple factorization. Root finding shows that it bifurcates into existence at a distance of $R = 2.33696\ldots$ with a value of $H_{MA}^2 = 5.32417\ldots$. Again, there is an inner and an outer solution. The inner solution continues down to a distance of $R = 3/2$, where the two groups touch and form an Euler configuration. The value of the angular momentum at this point equals 6.6125 and equals the value at which the Euler Rest Configuration terminates and the Euler Orbit Configuration is born. The outer solution continues its growth with increasing angular momentum. 

Now consider the energetic stability of these solutions. Similar to the Transverse Configurations, the mixed partials of the minimum energy function are identically zero at these relative equilibria. 
The second partials of Eqn.\ \ref{eq:delta_theta_M} with respect to $\theta$ yields
\beq
	\delta^2_{\theta\theta}{\cal E}_{AM} & = & \frac{R}{2} \left[ \frac{1}{(R-0.5)^3} - \frac{1}{(R+0.5)^3} \right]  (\delta\theta)^2
\eeq
which is always positive. The second partial of Eqn.\ \ref{eq:delta_R_M} with respect to $R$ is
\beq
	\delta^2_{RR}{\cal E}_{AM} & = & 2 \left[ \frac{9 H^2}{4(1.2+R^2)^3}\left(R^2-0.4\right) \right. \nonumber \\
	& & \left. - \frac{1}{(R-0.5)^3} - \frac{1}{(R+0.5)^3} \right] (\delta R)^2
\eeq
The resulting polynomial is of high order and is not analyzed. 
Alternately, inspecting the graph of this function shows that it crosses from negative to positive at the bifurcation point, as expected. Thus, initially the outer Aligned Configurations are energetically stable while the inner Aligned Configurations are unstable, and remain so until they terminate at the Euler configuration. To make a final check, evaluate the asymptotic sign of the second energy variation. For $R \gg 1$, $H^2_{MA} \sim 4/3R$. Substituting this into the above and allowing $R\gg 1$ again yields $\delta^2_{RR}{\cal E}_{AM} \sim 1/R^3 \delta R^2$, and thus the outer relative equilibria remain stable from their bifurcation on. 

The explicit energy of the Aligned Configurations are 
\beq
	{\cal E}_{MA} & = & \frac{3H^2}{4\left(1.2 + R^2\right)} - 1 - \frac{1}{(R-0.5)} - \frac{1}{(R+0.5)} 
\eeq
A direct comparison between ${\cal E}_{MA}$ and ${\cal E}_{MT}$ at the same levels of angular momentum shows that the Aligned Configurations always have a lower energy than the Transverse Configurations. This is wholly consistent with the energetic stability results found throughout.

\item
{\bf Purely Orbital Configurations}

Finally consider the purely orbital configurations for this case. As this is the sphere restricted problem, the orbital relative equilibria will be the same as exist for the point mass problem. 

\paragraph{Euler Solution}

For the Euler solution take the configuration where $r_{12} = r_{23} = R$ and $r_{31} = 2R$, $R \ge 1$, reducing the configuration to one degree of freedom. The minimum energy function then simplifies to
\beq
	{\cal E}_{OE} & = & \frac{3H^2}{2\left(6R^2 + 0.9\right)} - \frac{5}{2R}
\eeq
Taking the variation of the minimum energy function with respect to this configuration then yields
\beq
	\delta_R{\cal E}_{OE} & = & - \frac{18 H^2 R}{\left(6R^2+0.9\right)^2} + \frac{5}{2R^2} 
\eeq
Set this equal to zero and solving for the corresponding angular momentum to find
\beq
	H_{OE}^2 & = & \frac{5}{36} \frac{\left( 6R^2 + 0.9\right)^2}{R^3} 
\eeq
It can be shown that there are two orbital Euler configurations for $H_{OE}^2 > 8\sqrt{5}/3$ and none for lower values. The non-existence of solutions at a given total angular momentum occurs due to the coupling of the rotational angular momentum of the different bodies. In this case, however, the lower solutions all exist at $R <1$ and thus are not real for this system. In fact, given the constraint $R\ge 1$ there will be a single family of orbital Euler solutions at $H_{OE}^2 \ge 6.6125$ with corresponding radii ranging from $R = 1\rightarrow\infty$ as $H_{OE}^2 = 6.6125\rightarrow\infty$. 
The correspond energy of these Euler solutions as a function of $R$ is
\beq
	{\cal E}_{OE} & = & - \frac{5}{24R^3} \left(6R^2 - 0.9\right)
\eeq

Our simple derivation of the orbital Euler solutions only considers one-dimensional variations in the distance. However for a complete stability analysis it would be necessary to consider variations of each component in turn, as the instability in the Euler problem occurs transverse to the line of syzygies. However, the instability of the Euler solutions are well known and do not need to be derived here.

\paragraph{Lagrange Solution}

To find the conditions for the Lagrange solution take the configuration to be $r_{12} = r_{23} = r_{31} = R \ge 1$, again reducing the minimum energy function to a single degree of freedom.
\beq
	{\cal E}_{OL} & = & \frac{3H^2}{2\left(3R^2+0.9\right)} - \frac{3}{R} 
\eeq
The variation now yields the condition
\beq
	3 R \left[ \frac{1}{R^3} - \frac{3H^2}{\left(3R^2+0.9\right)^2}\right] & = & 0
\eeq
which can be solved for the angular momentum of the orbital Lagrange solutions as a function of orbit size
\beq
	H_{OL}^2 & = & \frac{\left(3R^2+0.9\right)^2}{3R^3} 
\eeq
Again, two solutions exist for $H_{OL}^2 > 16/\sqrt{10}$, however the inner solution has radius $R< 1$ and is not allowed by this model. Thus, again for the constraint $R\ge 1$ there is a single family of Lagrange solution orbits that range from $R = 1\rightarrow\infty$ as $H_{OL}^2 = 5.07\rightarrow\infty$. 
The corresponding energy of these Lagrange solutions as a function of $R$ is
\beq
	{\cal E}_{OL} & = & - \frac{1}{2R^3} \left(3R^2 - 0.9\right)
\eeq

Note again that the Lagrange solutions are always energetically unstable, and in particular that the equal mass Lagrange solutions are also spectrally unstable. As a final point, note that the energy of the Euler solutions is actually less than the energy of the Lagrange solutions when $R^2 < 63/60$, i.e., when $R$ is near unity. For larger values of $R$ the Lagrange solution is always lower energy. 
\end{enumerate}
\end{proof}

\begin{figure}[ht!]
\centering
\includegraphics[scale=0.28]{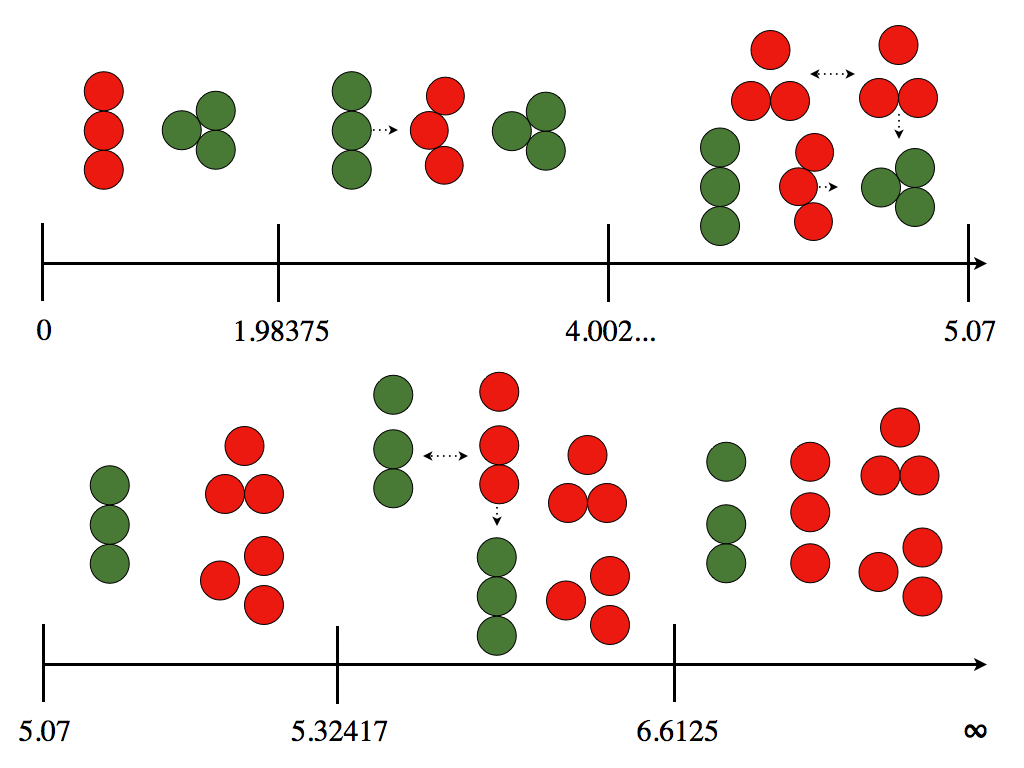}
\caption{Graphic depiction of all relative equilibria and their interactions. Green indicates stability while red indicates instability. The global minimum is not denoted in this diagram.}
\label{fig:N3bifurcation2}
\end{figure}

\begin{corollary}
The complete bifurcation chart of relative equilibria, minimum energy states, and global minimum energy states of the sphere restricted $N=3$ full body problem as a function of angular momentum follows. It is graphically illustrated in Figure \ref{fig:N3bifurcation2}
\begin{description}

\item
[$0 \le H^2 < 1.98375$] The Lagrange Rest Configuration is the only stable relative equilibria and is thus the minimum energy configuration. The Euler Rest Configuration does exist, but is unstable.

\item
[$H^2 = 1.98375$] The V Rest Configuration bifurcates into existence from the Euler Rest Configuration, and remains unstable throughout its life. The Euler Rest Configuration becomes marginally stable.

\item
[$1.98375 < H^2 < 2.99$] The Euler Rest Configuration becomes energetically stable, but the Lagrange Rest Configuration remains the global minimum energy configuration. 

\item
[$H^2 = 2.99$] The Euler and Lagrange Rest Configurations have equal energy and are both the global minimum energy configurations.

\item
[$2.99 < H^2 < 5.07$] The Euler Rest Configuration becomes the global minimum energy configuration, while the Lagrange Rest Configuration retains its energetic stability. 

\item
[$H^2 = 4.002\ldots$] The Transverse Mixed Configurations bifurcate into existence at a distance of $R=\sqrt{2.6}$.

\item
[$4.002\ldots < H^2 < 5.07$] The Transverse Mixed Configurations continue into an inner and outer family at higher values of angular momentum. Both of these families are energetically unstable throughout their life. The outer family exists for all higher values of angular momentum.

\item
[$H^2 = 5.07$] The V Rest and the Inner Transverse Mixed Configurations terminate at the Lagrange Rest Configuration. The Lagrange Rest Configuration ceases to exist and transitions into the Lagrange Orbital Configuration. 

\item
[$5.07 < H^2 < 5.32417\ldots$] The Euler Rest Configuration is the only stable relative equilibrium and is thus the global minimum energy configuration. The Lagrange Orbital Configuration exists for all higher angular momentum and is always unstable. 

\item
[$H^2 = 5.32417\ldots$] The Aligned Mixed Configurations bifurcate into existence at a distance of $R = 2.33696\ldots$. 

\item
[$5.32417\ldots < H^2 < 5.65907\ldots$] The Inner and Outer Aligned Mixed Configurations exist. The inner configuration is always energetically unstable throughout its life. The outer configuration is energetically stable and exists for all higher values of angular momentum. The Euler Rest Configuration remains the global minimum energy configuration. 

\item
[$H^2 = 5.65907\ldots$] The Euler Rest and Outer Aligned Mixed Configurations have equal energy and are both the global minimum energy configurations.

\item
[$5.65907\ldots < H^2 < 6.6125$] The Outer Aligned Mixed Configuration becomes the global minimum energy configuration, while the Euler Rest Configuration retains its energetic stability. 

\item
[$H^2 = 6.6125$] The Inner Aligned Mixed Configurations terminates at the Euler Rest Configuration. The Euler Rest Configuration ceases to exist and transitions into the Euler Orbital Configuration. 

\item
[$6.6125 < H^2$] The Outer Aligned Mixed Configuration is the only stable relative equilibrium and is thus the global minimum energy configuration, and remains so for all higher angular momentum. The Euler Orbital Configuration exists for all higher angular momentum and is always unstable. 

\end{description}

\end{corollary}

\begin{proof}
The transitions can all be evaluated by comparison between the existence and stability transitions outlined in the previous proof and by direct comparison of energy levels.
\end{proof}

\begin{figure}[ht!]
\centering
\includegraphics[scale=0.28]{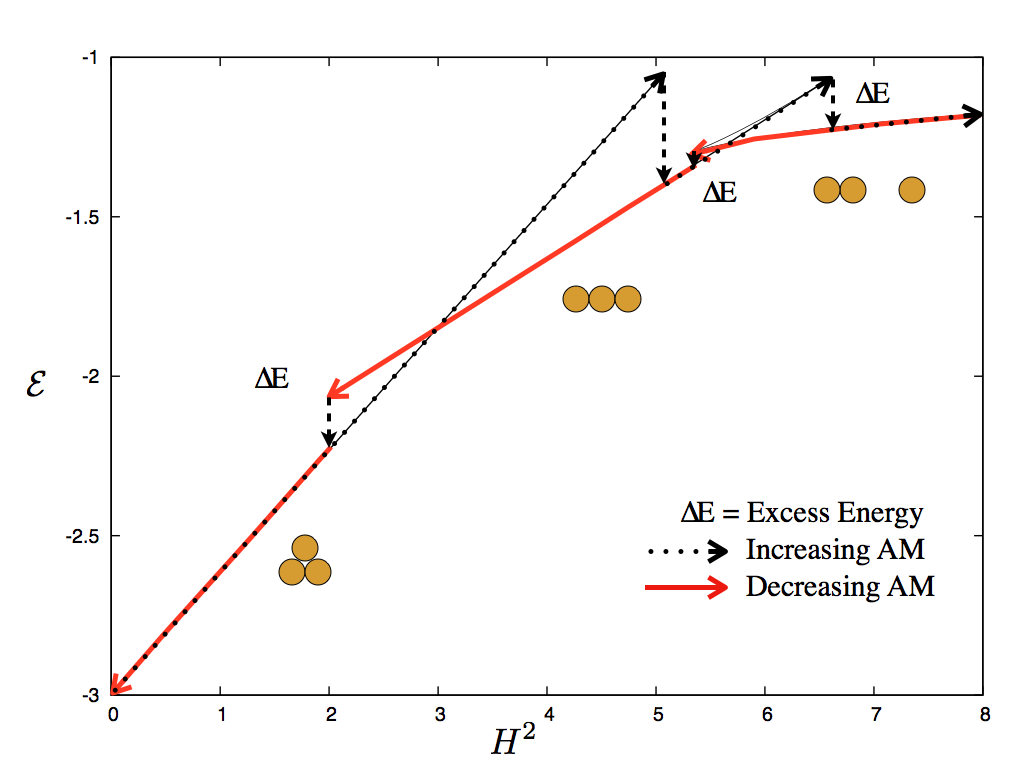}
\caption{Bifurcation diagram showing the energy-angular momentum curves of all stable relative equilibria and their transition paths for increasing and decreasing angular momentum.}
\label{fig:N3bifurcation}
\end{figure}

\begin{corollary}
The $N=3$ global and local minimum energy transition diagram as a function of angular momentum is discontinuous and exhibits hysteresis. 
Figure \ref{fig:N3bifurcation} tracks the energetically stable configurations as a function of angular momentum, shows where explicit transitions occur, and indicates the excess energy at each transition. To track the evolution of a system under changing angular momentum will require dynamical analysis at specific transition points.
\end{corollary}

\begin{proof}
The plot is created by comparing the energies of the different energetically stable configurations and computing energy differences at transitions. Since excess energies exist at different transition points, it will be necessary for the dynamical evolution of the system to be followed to determine actual outcomes.
\end{proof}

\subsection{$N\gg1$}

The paper ends with a few observations for when $N$ is a large number. Simulations of large numbers of grains that gravitationally attract each other and can rest on each other have been explored over the last decade using a variety of simulation methods. The original work in this area was performed by Richardson and collaborators (c.f.\ \cite{richardson_equilibrium}) and used hard-sphere models for surface interactions. More recent models using soft-sphere models for grain interactions have been explored by S\`anchez and Scheeres \cite{sanchez_ApJ}, and in some cases have advantages in terms of tracking the system-wide energy, angular momentum, and internal forces of these systems when there are significant components that rest on each other. Investigations of relative equilibria of the minimum energy function for such large particle simulations could be investigated, but have not been as of yet. 

There are several direct observations that can be made for large $N$ systems. First, for non-rotating systems it is expected that the global minimum configuration will tend towards a spherical collection of particles resting on each other. Depending on the presence and strength of friction between the grains, it is also possible that a variety of non-spherical shapes may be stable relative equilibria, thus generalizing the ideal results from continuum mechanics and theory. Likewise, as angular momentum is increased the classical Maclaurin and Jacobi ellipsoids should always be relative equilibria, however one must again note the probable existence of multiple other non-ellipsoidal configurations that are also relative equilibria. This is to be expected, given the presence of such multiple configurations at a given level of angular momentum for particle systems with $N > 2$. Of specific interest are the evolutionary pathways and transitions for these systems as the total angular momentum is increased (or decreased) mimicking the natural evolution of rubble pile asteroids under the YORP effect. Analytical studies of this have been made by Holsapple \cite{holsapple_yorp} and recreated using simulations by S\`anchez \cite{sanchez_icarus}. Of particular interest for these studies are the pathways to fission of large collections of particles. A systematic study of these transitions will be of real interest in understanding the overall evolution and creation of binary bodies among small asteroids, at the least\cite{jacobson_icarus}. 

To close this section, a final hypothesis is given. 
\begin{hypothesis}
For all $N\ge 2$, energetically stable relative equilibrium configurations of Full Body problems will result in one of two general states:
\begin{enumerate}
\item 
All the particles rest on each other in one collection and spin at a uniform rate; 
\item
All the particles separate into two collections in a mutually circular orbit about each other with doubly-synchronous rotation. 
\end{enumerate}
Further, that all configurations where the particles condensate into 3 or more collections will be energetically unstable and will either eject the excess collections or, under energy dissipation, will condense into one or two collections.
\end{hypothesis}

One possible avenue to the proof of this hypothesis is through the proof for the point mass $N$ body problem that all orbital relative equilibria with  $N\ge 3$ are energetically unstable. As for most collections with $N\ge3$ in relative equilibria, the leading order of their gravitational attractions will always represent the point mass $N$ body problem, with higher-order gravitational perturbations arising from the non-spherical configuration of the condensed grains in each collection.

\section{Conclusions}

This paper considers the basic and fundamental question: What are the minimum energy and stable configurations of the $N$-body problem in celestial mechanics for a given value of angular momentum. It is shown that this problem is ill-posed for the classical point mass model of celestial mechanics, but is well posed if all density distributions for these systems are made finite. This stipulation requires that rotational energy and angular momentum also be considered in searching for these stable configurations, and indeed fundamentally changes the problem. Specifically, under this reasonable stipulation the minimum energy solutions of the finite density 2-body problem change fundamentally as compared to the point mass 2-body problem. This question is also rigorously explored for $N=3$ along with a number of hypotheses for the $N\gg1$ problem. In the course of these discussions a number of fundamental questions are identified on the evolution of celestial mechanics systems under an increase of angular momentum, as has been documented to occur in nature. 

\section*{Acknowledgements}

Fruitful and motivating conversations with James Montaldi and Manuele Santoprete at a Mathematisches Forschungsinstitut Oberwolfach workshop in August 2011 are acknowledged. 
The author acknowledges support from NASA grants NNX11AP24G and NNX10AG53G from the Planetary Geology and Geophysics and the Near Earth Objects Observation programs, respectively. 

\newpage \bibliographystyle{plain}
\bibliography{../../../bibliographies/biblio_article,../../../bibliographies/biblio_books,../../../bibliographies/biblio_misc}

\begin{thebibliography}{10}

\bibitem{arnoldIII}
V.I. {Arnold}, V.V. {Kozlov}, and A.I. {Neishtadt}.
\newblock {\em {Mathematical aspects of classical and celestial mechanics}}.
\newblock Springer, 2006.

\bibitem{burns_safronov}
J.A. {Burns} and V.S. {Safronov}.
\newblock {Asteroid nutation angles}.
\newblock {\em Monthly Notices of the Royal Astronomical Society}, 165:403--+,
  1973.

\bibitem{castellanos}
A.~Castellanos.
\newblock The relationship between attractive interparticle forces and bulk
  behaviour in dry and uncharged fine powders.
\newblock {\em Advances in Physics}, 54(4):263--376, 2005.

\bibitem{cendra_marsden}
H.~Cendra and J.E. Marsden.
\newblock Geometric mechanics and the dynamics of asteroid pairs.
\newblock {\em Dynamical Systems, An International Journal}, 20:3--21, 2005.

\bibitem{fahnestock_icarus}
E.G. {Fahnestock} and D.J. {Scheeres}.
\newblock {Simulation and analysis of the dynamics of binary near-Earth
  Asteroid (66391) 1999 KW4}.
\newblock {\em Icarus}, 194:410--435, April 2008.

\bibitem{goldreich_peale}
P.~Goldreich and S.~Peale.
\newblock Spin-orbit coupling in the solar system.
\newblock {\em The Astronomical Journal}, 71:425, 1966.

\bibitem{goldreich}
P.~{Goldreich} and R.~{Sari}.
\newblock {Tidal Evolution of Rubble Piles}.
\newblock {\em Astrophysics Journal}, 691:54--60, January 2009.

\bibitem{holsapple_yorp}
K.A. Holsapple.
\newblock On yorp-induced spin deformations of asteroids.
\newblock {\em Icarus}, 205(2):430--442, 2010.

\bibitem{jacobson_icarus}
S.A. {Jacobson} and D.J. {Scheeres}.
\newblock {Dynamics of rotationally fissioned asteroids: Source of observed
  small asteroid systems}.
\newblock {\em Icarus}, 214:161--178, July 2011.

\bibitem{maciejewski}
A.J. Maciejewski.
\newblock Reduction, relative equilibria and potential in the two rigid bodies
  problem.
\newblock {\em Celestial Mechanics and Dynamical Astronomy}, 63(1):1--28, 1995.

\bibitem{moeckel_central}
R.~Moeckel.
\newblock On central configurations.
\newblock {\em Mathematische Zeitschrift}, 205(1):499--517, 1990.

\bibitem{pravec_fission}
P.~Pravec, D.~Vokrouhlick{\`y}, D.~Polishook, D.J.~Scheeres, A.W.~Harris,
  A.~Gal{\'a}d, O.~Vaduvescu, F.~Pozo, A.~Barr, P.~Longa, et~al.
\newblock Formation of asteroid pairs by rotational fission.
\newblock {\em Nature}, 466(7310):1085--1088, 2010.

\bibitem{richardson_equilibrium}
D.C. Richardson, P.~Elankumaran, and R.E. Sanderson.
\newblock Numerical experiments with rubble piles: equilibrium shapes and
  spins.
\newblock {\em Icarus}, 173(2):349--361, 2005.

\bibitem{saari_constant_polar}
D.G. Saari.
\newblock On bounded solutions of the n-body problem.
\newblock In {\em Periodic Orbits Stability and Resonances}, volume~1, page~76,
  1970.

\bibitem{sanchez_ApJ}
P.~S{\'a}nchez and D.J. Scheeres.
\newblock Simulating asteroid rubble piles with a self-gravitating soft-sphere
  distinct element method model.
\newblock {\em The Astrophysical Journal}, 727:120, 2011.

\bibitem{sanchez_icarus}
P.~S{\'a}nchez and D.J. Scheeres.
\newblock {DEM} simulation of rotation-induced reshaping and disruption of
  rubble-pile asteroids.
\newblock {\em Icarus}, 218: 876-894, 2012. http://dx.doi.org/10.1016/j.icarus.2012.01.014.

\bibitem{scheeres_F2BP}
D.J. Scheeres.
\newblock Stability in the full two-body problem.
\newblock {\em Celestial Mechanics and Dynamical Astronomy}, 83(1):155--169,
  2002.

\bibitem{scheeres_gravgrad}
D.J. {Scheeres}.
\newblock {Relative Equilibria for General Gravity Fields in the
  Sphere-Restricted Full 2-Body Problem}.
\newblock {\em Celestial Mechanics and Dynamical Astronomy}, 94:317--349, March
  2006.

\bibitem{scheeres_fission}
D.J. Scheeres.
\newblock Rotational fission of contact binary asteroids.
\newblock {\em Icarus}, 189(2):370--385, 2007.

\bibitem{scheeres_F2BP_planar}
D.J. Scheeres.
\newblock Stability of the planar full 2-body problem.
\newblock {\em Celestial Mechanics and Dynamical Astronomy}, 104(1):103--128,
  2009.

\bibitem{simo_marsden}
J.C. Simo, D.~Lewis, and J.E. Marsden.
\newblock Stability of relative equilibria. part i: The reduced energy-momentum
  method.
\newblock {\em Archive for Rational Mechanics and Analysis}, 115(1):15--59,
  1991.

\bibitem{smaleI}
S.~Smale.
\newblock Topology and mechanics. i.
\newblock {\em Inventiones mathematicae}, 10(4):305--331, 1970.

\bibitem{smaleII}
S.~Smale.
\newblock Topology and mechanics. ii.
\newblock {\em Inventiones mathematicae}, 11(1):45--64, 1970.

\bibitem{wang}
L.S. Wang, P.S. Krishnaprasad, and JH~Maddocks.
\newblock Hamiltonian dynamics of a rigid body in a central gravitational
  field.
\newblock {\em Celestial Mechanics and Dynamical Astronomy}, 50(4):349--386,
  1990.

\end{thebibliography}

\end{document}